\documentclass{revtex4-1}
\usepackage{latexsym, color, graphicx, comment}
\usepackage{amssymb}
\usepackage{amsmath}
\usepackage{hyperref}
\hypersetup{colorlinks=true,linkcolor=blue}

\newcommand{\vect}[1]{\mbox{\boldmath $#1$}}
\DeclareMathAlphabet{\mathbfsf}{\encodingdefault}{\sfdefault}{bx}{n}

\newcommand{\nescoil}{{\tt NESCOIL}}
\newcommand{\regcoil}{{\tt REGCOIL}}
\newcommand{\vmec}{{\tt VMEC}}
\newcommand{\currentPot}{\Phi}
\newcommand{\currentPotSV}{\Phi_{sv}}
\newcommand{\Bnormal}{B_{\mathrm{normal}}}
\newcommand{\sincos}{\begin{pmatrix} \sin \\ \cos \end{pmatrix}}
\newcommand{\cossin}{\begin{pmatrix} \cos \\ -\sin \end{pmatrix}}
\newcommand{\Bnplasma}{B_{\mathrm{normal}}^{\mathrm{plasma}}}
\newcommand{\Bnexternal}{B_{\mathrm{normal}}^{\mathrm{external}}}

\newcommand{\BnormalSV}{B_{\mathrm{normal}}^{\mathrm{sv}}}

\begin{document}

\title{An improved current potential method for fast computation of stellarator coil shapes}



\author{Matt Landreman}
\email[]{mattland@umd.edu}
\affiliation{Institute for Research in Electronics and Applied Physics, University of Maryland, College Park, MD, 20742, USA}


\date{\today}

\begin{abstract}

Several fast methods for computing stellarator coil shapes are compared,
including the classical NESCOIL procedure [Merkel, Nucl. Fusion 27, 867 (1987)],
its generalization using truncated singular value decomposition,
and a Tikhonov regularization approach we call REGCOIL in which the squared current density
is included in the objective function.
Considering W7-X and NCSX geometries,
and for any desired level of regularization,
we find the REGCOIL approach simultaneously achieves lower surface-averaged and maximum values of
both current density (on the coil winding surface) and normal magnetic field (on the desired plasma surface). 
This approach therefore can simultaneously improve the free-boundary reconstruction of the target plasma shape
while substantially increasing the minimum distances between coils,
preventing collisions between coils while
improving access for ports and maintenance.
The REGCOIL method also allows finer control over the level of regularization,
it preserves convexity to ensure the local optimum found is the global optimum,
and it eliminates two pathologies of  NESCOIL:
the resulting coil shapes become independent of the arbitrary choice of angles used to parameterize
the coil surface, and
the resulting coil shapes converge rather than diverge as Fourier resolution is increased.
We therefore contend that REGCOIL should be used instead of NESCOIL for applications in
which a fast and robust method for coil calculation is needed, such as when targeting coil complexity in
fixed-boundary plasma optimization, or for scoping new stellarator geometries.

\end{abstract}


\maketitle 

\section{Introduction}

Stellarators offer the promise of highly stable, disruption-free, steady-state fusion plasmas with no power recirculated for current drive, but there are also significant challenges for the stellarator concept related to the
complicated magnetic field coils. The magnetic field coils are among
the most expensive components in any magnetic fusion system, and the non-planar coils of a stellarator are particularly 
difficult to design and fabricate due to the need for precise three-dimensional shaping. 
The ability of non-planar coils to shape the magnetic field is constrained by the requirement that the coils cannot collide (overlap).
Even without collisions, the minimum coil-to-coil separation in stellarators is typically small, limiting
access between the coils for diagnostics and heating systems, and for blanket maintenance in a reactor.
Any advances in stellarator coil design can potentially have a significant impact on the cost and feasibility of fusion energy.

For recent optimized stellarators such as W7-X \cite{Greiger,Klinger} and HSX \cite{HSX}, the coil design has been part of a two-stage optimization process.
In the first stage, the plasma boundary shape is varied to optimize physics properties such as neoclassical transport and magnetohydrodynamic (MHD) stability, with a fixed-boundary solution of the MHD equilibrium equations at each iteration, and no specific coil shapes computed. In the second stage, coil shapes are calculated that approximately produce the `target' plasma shape resulting from stage 1. This two-stage approach is advantageous both because it is computationally efficient (there is no need to compute free-boundary equilibria from a new coilset at each iteration in stage 1),
and because good flux surface quality is more likely than in a single-stage variation of coil shapes.

For this multi-stage optimization procedure, a central tool of stellarator coil design has been the \nescoil~code \cite{merkel_nescoil}.
In the \nescoil~approach, one first defines a coil winding surface on which the coils will lie, 
which might be done by expanding the target plasma boundary outward by some uniform or nonuniform distance. One then considers a `current potential' $\currentPot$, which is a scalar stream function for a divergence-free surface current $\vect{K}$ flowing on this coil surface:
\begin{equation}
\vect{K} = \vect{n}\times\nabla \currentPot,
\label{eq:currentPot}
\end{equation}
where $\vect{n}$ is the unit surface normal. \nescoil~computes the current potential that best produces the target plasma shape, in a least-squares sense.
Remarkably, this problem can be solved by solving a single linear system of modest size (typically $\le 144\times 144$),
making the procedure fast and robust. 
(The typical time for a well-resolved calculation, which is dominated by matrix assembly, is on the order of a few seconds on a modern CPU.)
Once the continuous current potential is obtained, the shapes of discrete coils are defined by taking a finite number of contours of $\currentPot$.
There is essentially only one parameter requiring routine adjustment, the number of Fourier modes retained in $\currentPot$,
making the code minimally complicated to use. 
On the other hand, \nescoil~does not permit the direct inclusion of other important optimization criteria or constraints,
such as the maximum magnetic field at the coils, minimum coil radius of curvature, or minimum separation between coils.
For this reason, nonlinear optimization tools for stellarator coil design have been developed, including {\ttfamily ONSET}\cite{onset},
{\ttfamily COILOPT}\cite{coilopt,coilopt2} (used to design NCSX \cite{Zarnstorff}), and {\ttfamily COILOPT++}\cite{Brown}.

While it is certain that such nonlinear optimization with engineering constraints will be essential to the design of future stellarator
facilities,
the \nescoil~approach remains widely used, for several reasons. First, due to \nescoil's speed, robustness, and relatively minimal complexity, it is often used for initial studies of new stellarator configurations, before a detailed engineering design is conducted.
Recent examples of such use include \cite{ku_QA,ku_QHS,estelle,QAS-LA,ust1}.
Second, the convergence of a nonlinear coil optimizer is expedited by using a good initial guess for the coil shapes, and \nescoil~has
been used to provide this initial condition \cite{estelle}.
Third, \nescoil~is sometimes called in the first stage of plasma optimization, to guide the fixed-boundary plasma optimization
towards configurations consistent with realistic coils \cite{pomphrey}.
Since the maximum feasible coil-to-plasma distance (or equivalently the coil complexity at fixed coil-to-plasma distance) 
is a strong function of the plasma shape \cite{LandremanBoozer},
this application is highly important. \nescoil's speed and reliability make it well suited for use inside the phase-1 optimization iterations in this manner. 
For all these applications, \nescoil~is uniquely robust in part because the optimization problem it solves is convex, so
\nescoil~can never be `stuck' in a local minimum that is not the global minimum, unlike most other coil optimization methods.

In this paper, we compare \nescoil~to two variants of the original method. The first variant is a truncated singular value decomposition (TSVD) option that was added to the \nescoil~code during the NCSX design process \cite{valanjuPoster,pomphrey}.
In the second variant, which we call \regcoil~(regularized \nescoil) for brevity, the average squared current density is added as an optimization criterion, in a manner that preserves the linearity, convexity, and speed of the original \nescoil~method.
We will demonstrate that for both W7-X and NCSX geometries, the \regcoil~approach systematically yields 
better coil shapes than the other two methods.
This is shown first in the sense that the maximum and average-squared current density is lower for a given level of error in the magnetic
field on the plasma boundary.
(Or equivalently, for any given maximum or average-squared current density, the error in the magnetic field
is lower.)
We also demonstrate these advantages using free-boundary reconstructions with discrete coils,
showing coilsets from \regcoil~can more accurately reconstruct the target plasma shape in practice than
coilsets from the other methods, while simultaneously having greater coil-to-coil separation.
Furthermore, we point out a shortcoming of the original \nescoil~and TSVD methods,
that the resulting coil shapes can be sensitive to the arbitrary choice of how the coil surface is parameterized,
meaning the code user must pay attention to this parameterization,
and potentially causing problems with use of \nescoil~inside an optimizer.
In contrast, the \regcoil~formulation yields results
that are independent of the choice of parameterization. 
Thus, we will show that \regcoil~has multiple advantages compared to the original \nescoil~and TSVD variants,
at comparable computational cost.

In most coil optimization work to date, 
departures from the target magnetic field  
are weighted uniformly in the objective function (in a sense
that will be made precise in section \ref{sec:nescoil}.)
However it would be preferable to place greater weight on the magnetic field errors that couple strongly to important physics properties,
such as magnetic field errors with a helicity that resonates with the rotational transform somewhere in the plasma,
driving magnetic islands.
Methods have been proposed for incorporating this plasma sensitivity information into coil design \cite{bnpert, BoozerJPP}.
Here, for simplicity, we take the usual approach of uniform weighting of errors, to focus on the comparison of regularization methods.

\section{Ill-posed problems and regularization}

\begin{figure}[h!]
\includegraphics[width=3in]{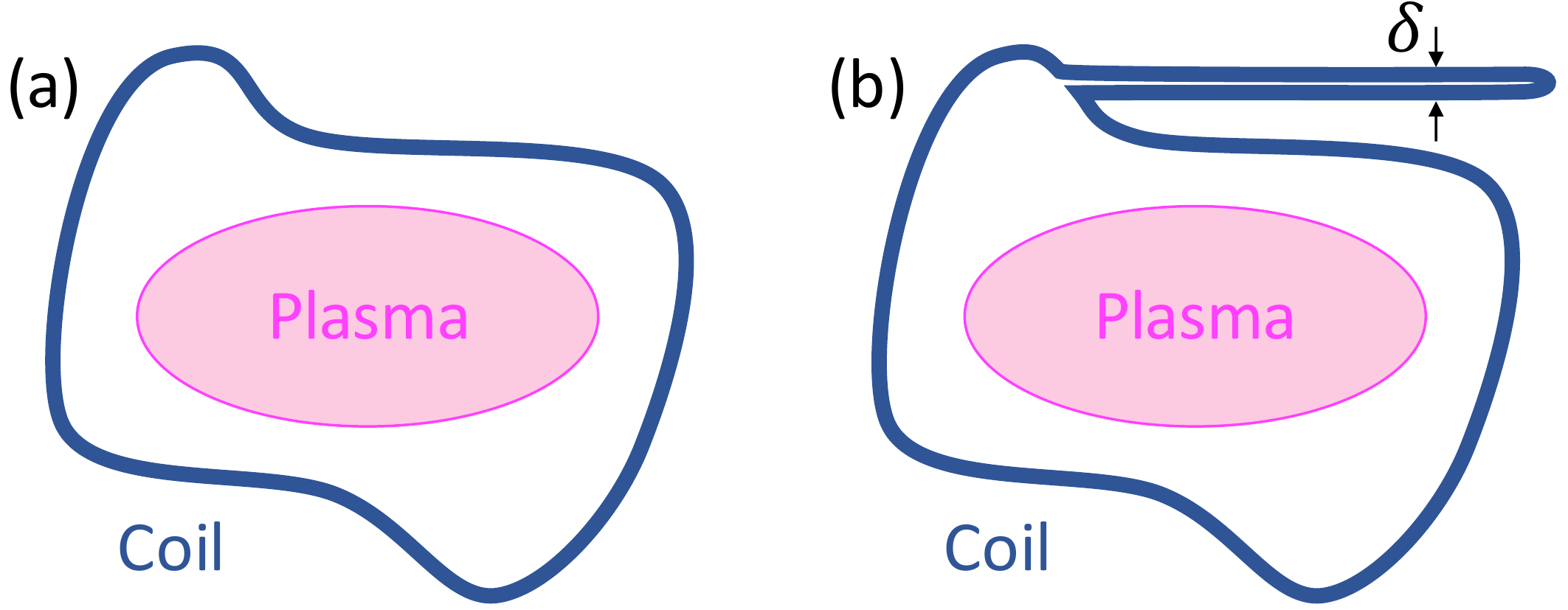}
\caption{(Color online)
Finding coil shapes that produce a given magnetic field is an ill-posed problem
since $O(1)$ changes to the coil shape,
such as the change from (a) to (b) here, can produce negligible change to the magnetic field
in the region of interest. Here, the difference between (a) and (b) of the magnetic field in the plasma region becomes
arbitrarily small as the distance $\delta \to 0$ due to near cancellation in the Biot-Savart law. 
\label{fig:illPosedIllustration}}
\end{figure}

Before describing the three algorithms in detail, one should note an important fact:
the problem of finding currents that produce a given plasma shape
is fundamentally ill-posed. Since the magnetic field from a coil decays with distance,
and since oppositely directed currents can give nearly cancelling contributions to the Biot-Savart integral,
certain aspects
of the coil shapes are effectively undetermined.
Given one set of coils that produce a certain plasma shape, one can in principle always add an
out-and-back segment to any coil without significantly affecting the plasma shape, as in figure \ref{fig:illPosedIllustration}.
Thus, there 
cannot be a unique solution for the coil shapes, unless extra criteria are provided
to disallow configurations like that of figure \ref{fig:illPosedIllustration}.b.
Analogous ill-posed problems arise in many other fields \cite{HansenBook}, 
often in the context of least-squares problems, another example from fusion being tomography \cite{Jacobsen}.

In any numerical scheme that attempts to solve a fundamentally ill-posed problem,
typically a linear system will arise for which the matrix has a large condition number \cite{HansenBook}, with a solution that does not converge as numerical resolution is increased.
Ill-posed problems can be `regularized' in a variety of ways by imposing some additional constraints,
such that the underlying problem becomes well-posed, and the solution is no longer underdetermined.
One regularization approach can be to look for the solution only within a particular subspace, such as the space of
polynomials of degree $p$ for some finite $p$. 
Another regularization approach can be to impose a constraint on the solution norm.
The different algorithms considered below for computation of stellarator coil shapes essentially differ in their choice of regularization technique.

In any regularization method for an ill-posed problem, one is always left with a degree of arbitrariness in the solution,
represented by some parameter that reflects how much regularization to impose. 
As the original problem has an infinite number of solutions, 
 this arbitrariness in the solution is unavoidable. 
The solution to any regularized ill-posed problem typically transitions from smooth to noisy (or vice-versa) 
as this regularization parameter is varied.
This behavior is familiar to anyone who has fit data with a polynomial, in which case the regularization parameter
is the polynomial degree $p$: raising $p$ reduces differences between the data and fit, but if $p$ is too large there is overfitting.
There is no universal rule to choose the regularization parameter; the parameter often must 
be chosen by trial-and-error, or by using additional information or constraints outside the problem formulation.
Indeed, we will find that the three coil-computation algorithms each has one regularization parameter
that the user is forced to choose.

The existence of an arbitrary regularization parameter also reflects the fact that coil design
is an optimization problem with multiple competing objectives. The coils should produce a plasma with the desired shape
(or more to the point, a plasma with good physics properties),
but the coils should also be far from each other, with large radius of curvature, etc. 
A multi-objective optimization problem  \cite{steuer} generally does not have a unique solution, 
in the sense that an improvement can typically be achieved for one objective 
if a sacrifice is made in another objective.
A unique solution only exists if some additional information is supplied to weigh the competing objectives.
The regularization parameter is precisely this additional information,
a relative weight given to the coil shape complexity compared to the plasma shape reconstruction.

\section{Linear current potential methods for computing stellarator coil shapes}

\subsection{NESCOIL}
\label{sec:nescoil}

In the \nescoil~method, an objective function $\chi^2_B$ is defined in terms of $\Bnormal$,
the normal component of the magnetic field on the desired (target) plasma surface,
by integrating its square over the plasma surface:
\begin{equation}
\chi^2_B = \int d^2a\; B^2_{\mathrm{normal}}.
\label{eq:chi2B}
\end{equation}
The fixed-boundary target plasma configuration has $\Bnormal=0$ exactly, but any attempt to produce this desired plasma
shape using external coils will result in some finite error $\Bnormal \ne 0$, leading to $\chi^2_B>0$. 

Next, note that the current potential $\currentPot$ is generally not single-valued, but by integrating (\ref{eq:currentPot})
around the coil surface poloidally or toroidally it follows that
\begin{equation}
\currentPot (\theta', \zeta') = \currentPotSV (\theta', \zeta')+ \frac{G \zeta'}{2\pi} + \frac{I \theta'}{2\pi},
\label{eq:currentPotDecomp}
\end{equation}
where $\currentPotSV$ is single-valued, $\theta'$ and $\zeta'$ are any poloidal and toroidal angles on the coil surface,
and $G$ and $I$ are the currents linking the coil surface poloidally and toroidally.
Here and throughout, we use primes to indicate coordinates on the coil surface, with unprimed coordinates indicating the plasma surface.
The secular terms in (\ref{eq:currentPotDecomp}) involving $G$ and $I$ are considered known, since $G$ and $I$
are determined by the desired coil topology and target plasma configuration.
(For example, saddle coils are defined by $G=I=0$ so the coil filaments do not link the coil surface. For modular coils, $I=0$, and $G$ can be determined
from the target plasma equilibrium by applying Ampere's Law to a toroidal loop at the plasma edge.)
The goal becomes determination of the unknown single-valued term $\currentPotSV$. 

The linearity of Ampere's Law and (\ref{eq:currentPot})
mean that $\Bnormal$ can be written as a superposition of the contributions from various currents:
\begin{equation}
\Bnormal = \Bnplasma + \Bnexternal + \Bnormal^{GI} + \BnormalSV\{ \currentPotSV \},
\end{equation}
where $\Bnplasma$ is the component arising due to currents in the plasma,
$\Bnormal^{GI}$ is the component arising due to the secular $G$ and $I$ terms in the current potential (\ref{eq:currentPotDecomp}),
$\BnormalSV$ is the component arising due to the single-valued term  $\currentPotSV$ in (\ref{eq:currentPotDecomp}),
and $\Bnexternal$ is the component arising to due to any other external coils that might exist, such as planar toroidal field coils.
The operator $\BnormalSV$ can be derived from the Biot-Savart law, so it is linear,
and an explicit formula is given in appendix \ref{sec:equations}. 
In essence, one wants to determine the $\currentPotSV$ 
that approximately drives an equal and opposite $\Bnormal$ to the other terms in $\Bnormal$,
$\Bnplasma + \Bnexternal + \Bnormal^{GI}$.

The optimization problem is discretized by writing $\currentPotSV$ as a finite Fourier series with amplitudes $\currentPot_j$:
\begin{equation}
\currentPotSV(\theta',\zeta') = \sum_j \currentPot_j \sincos_j (m_j \theta' - n_j \zeta'),
\label{eq:Fourier}
\end{equation}
where $j$ indexes both the poloidal and toroidal mode numbers $(m,n)$ as well as indicating whether
a sine or cosine phase is used. For stellarator-symmetric geometries, only the sine phase is needed.

Determination of the $\currentPot_j$ that minimizes $\chi^2_B$ then has the form of a
linear least-squares problem. 
Detailed expressions for the method are given in appendix \ref{sec:equations}.
As with any linear least-squares problem, the solution can be obtained using the normal equations, $QR$ decomposition,
or singular value decomposition.
The number of unknowns ($\currentPot_j$) retained is typically 24-144, i.e. quite modest.	
Thus, the procedure is robust and fast: the current potential 
is computed by solving a single linear system, with no iteration required.
The solution is unique, so there is no danger of getting trapped in a local minimum that is
not the global minimum, as could be the case for non-convex optimization problems.

In the \nescoil~approach, the ill-posed problem is regularized through the truncation of the Fourier expansion (\ref{eq:Fourier}),
i.e. only optimizing the current potential within a restricted space of functions.
Typically modes $m_j$ and $n_j$ are included which satisfy $|m_j| \le N_F$ and $|n_j| \le N_F$
for some number $N_F$ in the range 3-8.

\subsection{Truncated SVD}

Any linear least-squares problem such as the one solved by \nescoil~may be solved using a singular value decomposition (SVD). 
The SVD of a $m\times n$ matrix $A$ is $A=U \Sigma V^T$ where $U$ is a $m\times m$ orthogonal matrix,
$V$ is a $n\times n$ orthogonal matrix, and $\Sigma$ is a $m\times n$ diagonal matrix,
with the diagonal entries $\sigma_j$ non-negative and in decreasing order. The columns of $U$ and $V$ are known
as the left and right singular vectors respectively, and the $\sigma_j$ are the singular values. For the linear least-squares problem
of determining the $n$-element vector $x$ that minimizes $|Ax-b|^2$ for a $m$-element vector $b$,
it can be shown that the solution is $x = V \Sigma^{+} U^T b$. Here $\Sigma^{+}$ is the $n\times m$ diagonal matrix with
$j$th diagonal element equal to $1/\sigma_j$ if $\sigma_j>0$, or to 0 if $\sigma_j=0$. 
The matrix $V \Sigma^{+} U^T$ used to compute the solution is called the pseudoinverse of $A$.
For \nescoil, $x$ corresponds to $\currentPot_j$,
and $A$ and $b$ are given in (\ref{eq:lsq}).

A standard approach to regularizing ill-posed least-squares problems is using a truncated SVD (TSVD) \cite{HansenBook}.
In this method, all diagonal elements of $\Sigma^{+}$ are replaced with 0 after row $\ell$,
where $\ell$ becomes the regularization parameter. 
That is, only $\ell$ singular values and singular vectors are retained in the pseudoinverse.
The idea behind this method is that
the decreasing sequence of singular values (with corresponding singular vectors) 
corresponds to patterns in the solution that contribute less and less to the residual.
Beyond a certain point in this sequence, these patterns in the solution are effectively undetermined,
so their amplitude is set to 0.
A TSVD solution to the minimization problem (\ref{eq:lsq}) was added to the \nescoil~code during the NCSX 
design process \cite{valanjuPoster,pomphrey}.

\subsection{REGCOIL}

Another widely-used technique for regularizing ill-posed least-squares problems in other fields
is Tikhonov regularization,
sometimes also called ridge regression \cite{Tikhonov1963,HansenBook}.
In Tikhonov regularization, the original least-squares problem
$\min_x\{|Ax-b|^2\}$ is augmented with a second quadratic term:
\begin{equation}
\min_x \{|Ax-b|^2 + \lambda | L(x-x_0)|^2 \},
\label{eq:Tikhonov}
\end{equation}
for some scalar $\lambda$, matrix $L$, and vector $x_0$. 
The idea is to regularize the problem by finding a solution $x$ for which the difference from $x_0$
has a small norm, in a sense given by $L$. In the common case that $x_0=0$ and $L$ is the identity matrix,
it is just the usual 2-norm of the solution that is minimized. The regularization parameter is $\lambda$,
representing how much importance to place on minimizing the solution norm
compared to the competing goal of minimizing the original residual.

The \regcoil~method for computing stellarator coil shapes
is the application of Tikhonov regularization to the least-squares problem from \nescoil,
taking the regularization term $|L(x-x_0)|^2$ to be 
a physically meaningful quantity: the surface-average-squared current density:
\begin{equation}
\chi^2_K = \int d^2a' \; K(\theta',\zeta')^2,
\label{eq:chi2K}
\end{equation}
where $K=|\vect{K}|$ reflects the inverse distance between coils.
An explicit expression for $\chi^2_K$ in terms
of the unknown $\currentPot_j$ is given in appendix \ref{sec:equations}, where it can be seen that 
$\chi^2_K$ indeed has the form of the regularization term in (\ref{eq:Tikhonov}).
In other words, \regcoil~finds the current potential that minimizes the objective function $\chi^2$ defined by
\begin{equation}
\chi^2 = \chi^2_B + \lambda \chi^2_K,
\label{eq:chi2tot}
\end{equation}
with regularization parameter $\lambda$.
Crucially, since $K^2$ is a quadratic function of $\currentPotSV$,
the problem of finding the $\currentPotSV$ that minimizes (\ref{eq:chi2tot})
remains a \emph{linear} least-squares problem,
with a unique solution that can be obtained rapidly using normal equations, $QR$ decomposition, or a SVD.
Other constraints that one might include in a detailed engineering optimization, such as
the maximum $K$, radius of curvature of the coils, or maximum magnetic field strength $B$ on the coils,
would require an iterative optimization, and would potentially introduce the possibility of multiple local minima
in the objective function.

Although the \regcoil~method had been implemented in a U.S. version of the \nescoil~code previously,
we are unaware of any publications in which the method has been described or demonstrated.
Results shown below were generated using an independent parallelized implementation of \regcoil, using {\ttfamily OpenMP} parallelization for assembling the matrices, (typically the most expensive step in any of these three methods),
and using multi-threaded {\ttfamily LAPACK} routines for solving the least-squares problem.

\section{Variation with level of regularization}

\begin{figure}[h!]
\includegraphics[width=3in]{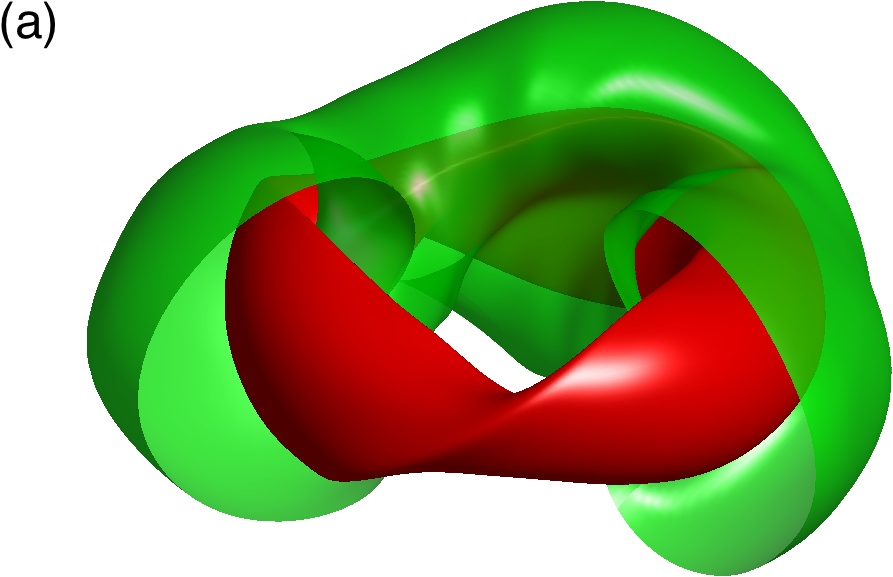}
\includegraphics[width=3in]{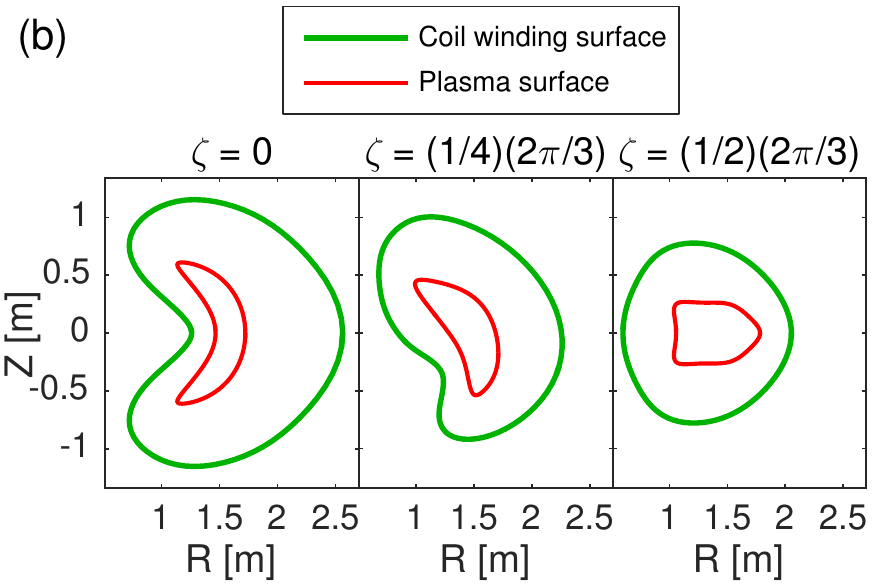}
\includegraphics[width=3in]{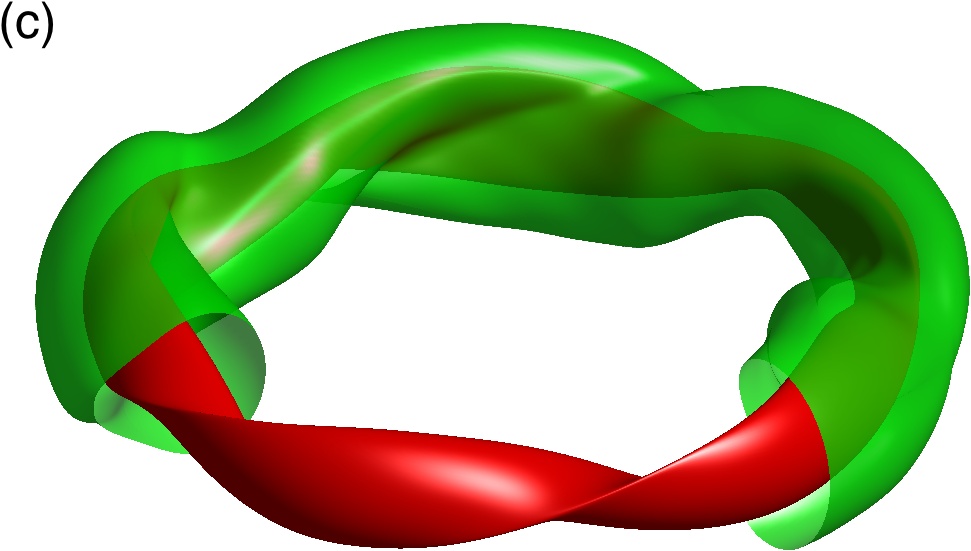}
\includegraphics[width=3in]{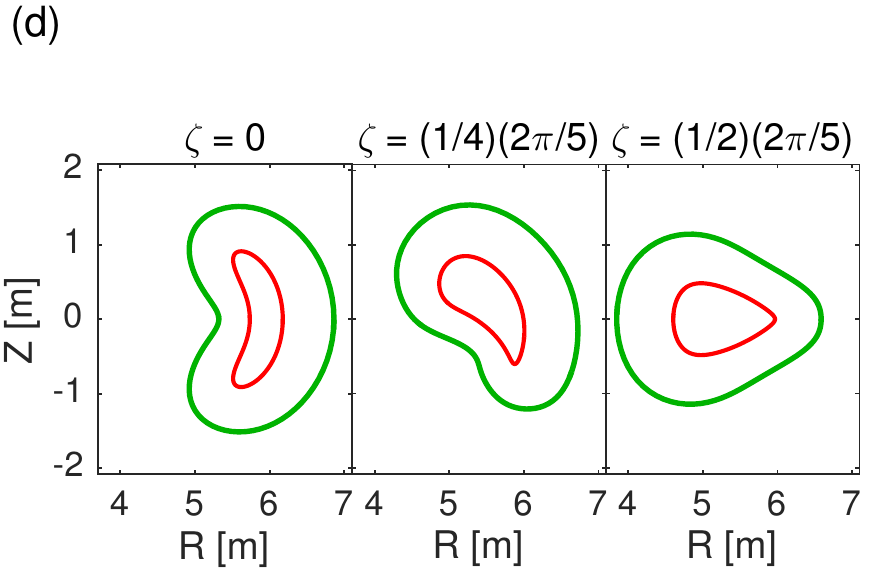}
\caption{(Color online)
Shapes of the plasma and coil winding surfaces used in this work.
(a)-(b) NCSX
(c)-(d) W7-X. 
\label{fig:geometry}}
\end{figure}

We will apply the three coil calculation algorithms to two geometries, NCSX \cite{Zarnstorff} and W7-X \cite{Greiger,Klinger},
shown in figure \ref{fig:geometry}. In both cases the plasma surface is taken from a fixed-boundary
\vmec~equilibrium \cite{VMEC1983} free of coil ripple. For NCSX this equilibrium is the one known as LI383.
For each geometry, the coil surfaces used are surfaces on which the actual experimental coils lie. 

\begin{figure}[h!]
\includegraphics[width=6.5in]{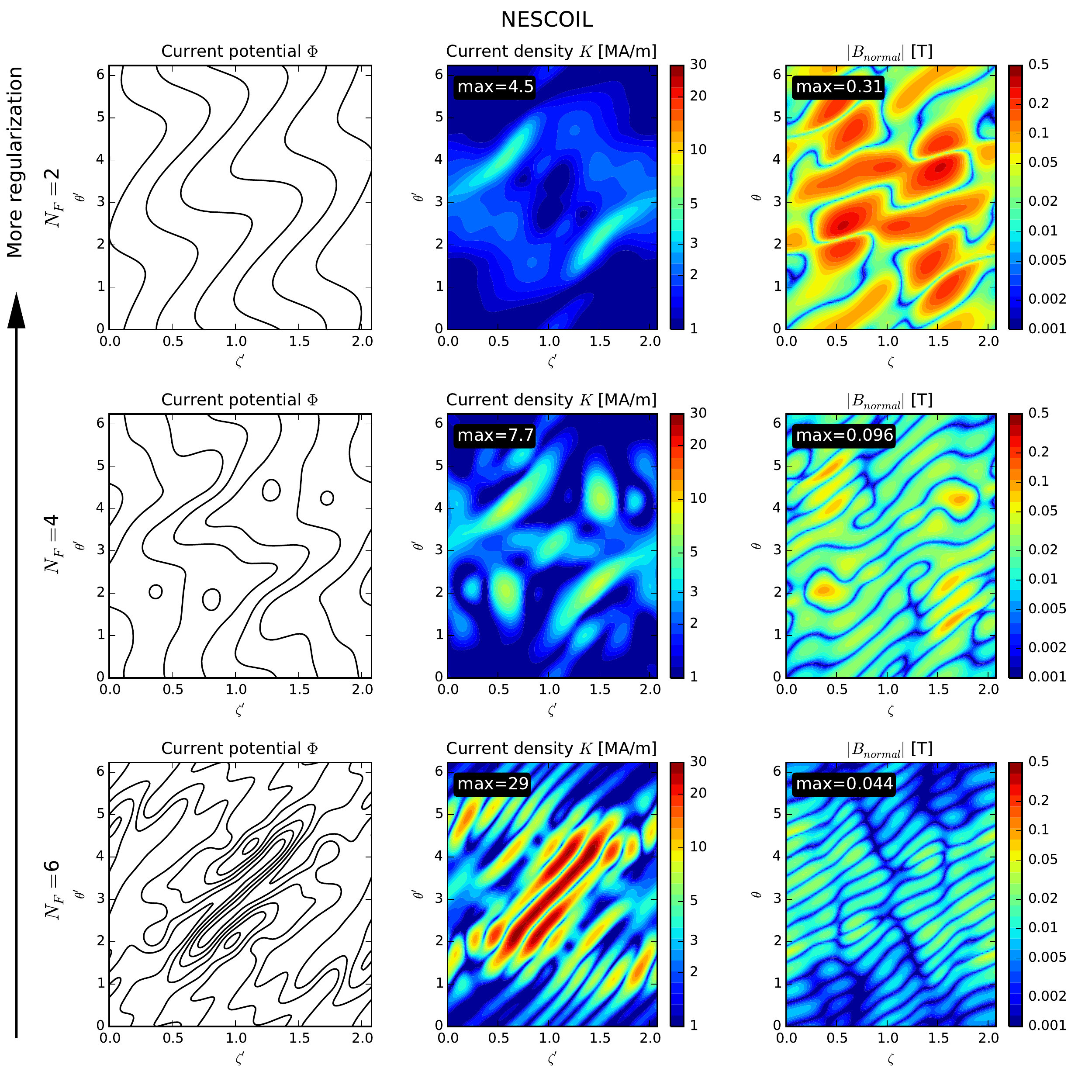}
\caption{(Color online)
Application of  \nescoil~to the NCSX geometry, showing trends with varying regularization parameter $N_F$. 
Regularization is provided in \nescoil~by truncating the Fourier expansion of the current potential to modes with $|m| \le N_F$ and
$|n| \le N_F$, so larger $N_F$ corresponds to weaker regularization.
As the level of regularization is increased,
the coil shapes given by contours of the current potential become less complicated,
and the current density $K$ on the coil surface decreases, but the residual $|\Bnormal|$ on the target plasma surface increases.
Poloidal and toroidal angles on the coil surface are $(\theta',\zeta')$, and corresponding angles on the plasma surface are $(\theta,\zeta)$.
\label{fig:decreasingRegularization_nescoil}}
\end{figure}

\begin{figure}[h!]
\includegraphics[width=6.5in]{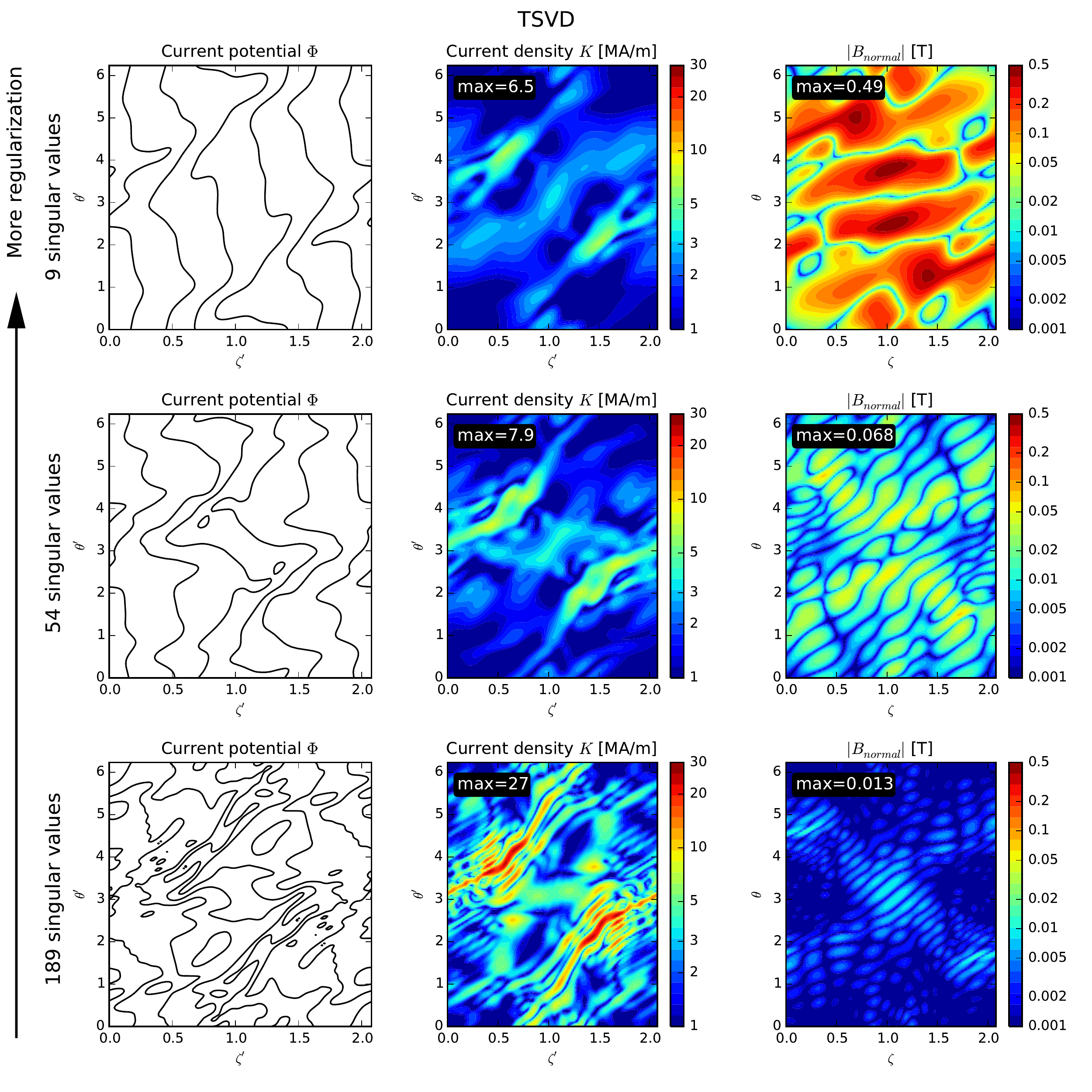}
\caption{(Color online)
Application of  the truncated singular value decomposition (TSVD)
method to the NCSX geometry, showing trends with varying regularization. 
Regularization is provided in the TSVD approach by truncating the number of singular values retained in the pseudoinverse,
so more singular values corresponds to less regularization.
As the level of regularization is increased,
the coil shapes given by contours of the current potential become less complicated,
and the current density $K$ on the coil surface decreases, but the residual $|\Bnormal|$ on the target plasma surface increases.
\label{fig:decreasingRegularization_tsvd}}
\end{figure}

\begin{figure}[h!]
\includegraphics[width=6.5in]{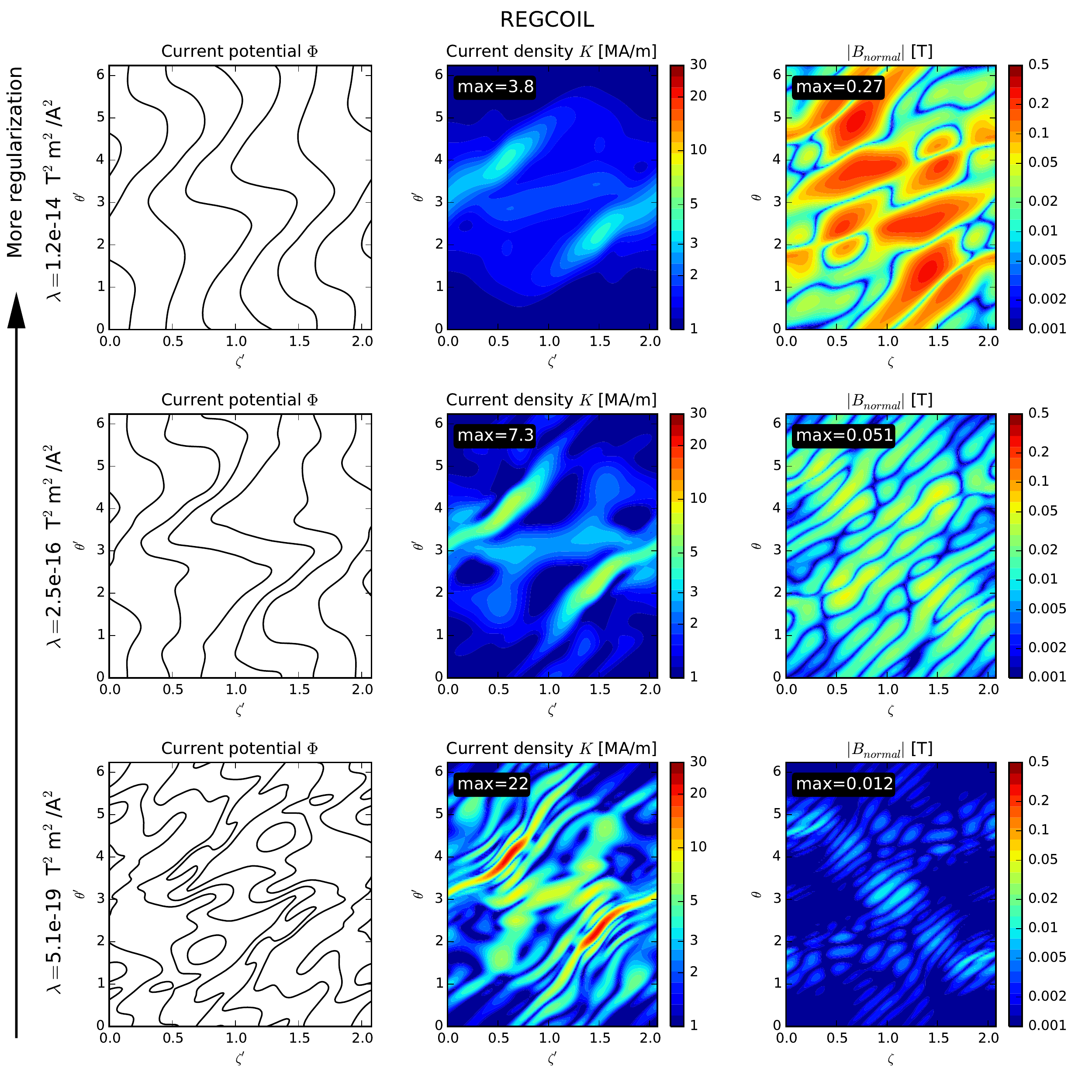}
\caption{(Color online)
Application of  \regcoil~to the NCSX geometry, showing trends with varying regularization parameter $\lambda$. 
As the level of regularization is increased,
the coil shapes given by contours of the current potential become less complicated,
and the current density $K$ on the coil surface decreases, but the residual $|\Bnormal|$ on the target plasma surface increases.
\label{fig:decreasingRegularization_regcoil}}
\end{figure}

As a first application, we illustrate how the three methods behave as the level of regularization is varied in each approach.
While the specific mechanism of regularization is different in the three methods, several trends are similar:
as the level of regularization is increased, the coil shapes become less complicated, the current density $K$ decreases, and 
the residual $|\Bnormal|$ increases. Figures \ref{fig:decreasingRegularization_nescoil}, \ref{fig:decreasingRegularization_tsvd}, 
and \ref{fig:decreasingRegularization_regcoil}  demonstrate these trends for the NCSX geometry
for the three algorithms. For \nescoil, the level of regularization is increased by lowering $N_F$, the maximum poloidal
and toroidal Fourier mode number. For the TSVD method, the level of regularization is increased by including fewer singular values 
(with corresponding singular vectors) in the pseudoinverse. For \regcoil, the level of regularization is increased by increasing $\lambda$.
The qualitative trends in the three methods are similar. In particular, $K$ decreases as $|\Bnormal|$ increases. 
Correlated with decreasing $K$ is decreasing complexity of the coil shapes given by the contours
of the current potential.
Ideally one wants both low $K$ and low $|\Bnormal|$, but there is a tradeoff between these competing objectives for
all three algorithms. 

\begin{figure}[h!]
\includegraphics[width=3in]{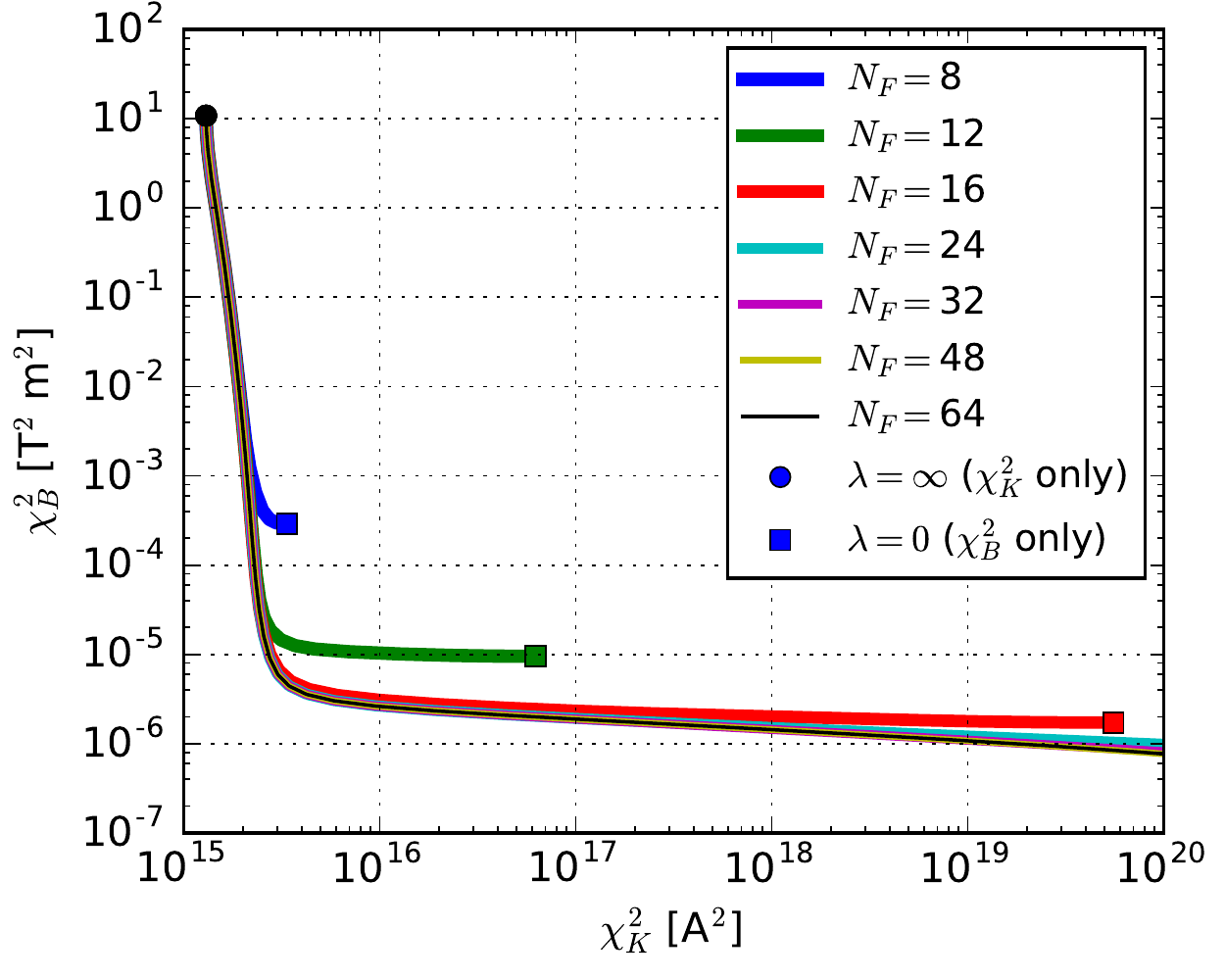}
\caption{(Color online)
Convergence of the \regcoil~method, using W7-X geometry. As the Fourier resolution $N_F$ is increased,
more and more of the tradeoff curve is resolved,
corresponding to solutions of lower regularity.
\label{fig:convergence}}
\end{figure}

While a finite Fourier series is used for the current potential in all three methods,
increasing the Fourier resolution $N_F$ causes the current potential from \nescoil~to diverge,
whereas results for the TSVD and \regcoil~methods converge. The divergence of \nescoil~can
be seen in figure \ref{fig:decreasingRegularization_nescoil}, while the convergence of \regcoil~is
demonstrated in figure \ref{fig:convergence}. In this figure, the tradeoff curve of $\chi^2_B$ versus $\chi^2_K$
is plotted for the full range of $\lambda$ from 0 to $\infty$, varying $N_F$. More and more of the curve
can be resolved as $N_F$ is increased. 
The $\lambda=0$ end of the curve corresponds to
\nescoil, so it diverges to $\chi^2_K \to \infty$ as $N_F \to \infty$. For any finite positive $\lambda$,
if the Fourier resolution $N_F$ is low enough, more regularization is provided by the Fourier truncation than by
the physical $\chi^2_K$ regularization, leading to higher $\chi^2_B$ than the converged values; such behavior
can be seen in  figure \ref{fig:convergence} for the $N_F=$8, 12, and 16 curves. In practice, the interesting solutions lie on the vertical
part of the L-shaped curve, for $\chi^2_B > 10^{-4}$ T$^2$m$^2$. Thus, we find $N_F \sim 12$ is typically plenty for
well converged results. Convergence behavior of the TSVD method is analogous to \regcoil.
For results shown throughout this paper, 
for the surface integrations detailed in  appendix \ref{sec:equations},
grid resolutions of $128\times128$ points on the plasma and coil surfaces were used for all three
methods, and it was verified that differences were negligible if a resolution of $256\times256$ was used.
Computing the $N_F=12$ curve of figure \ref{fig:convergence},
representing 100 unique coilsets with different values of $\lambda$,
required 7 seconds on 1 node of the NERSC Cori computer.

\section{Comparison of the methods}

\begin{figure}[h!]
\includegraphics[width=6.5in]{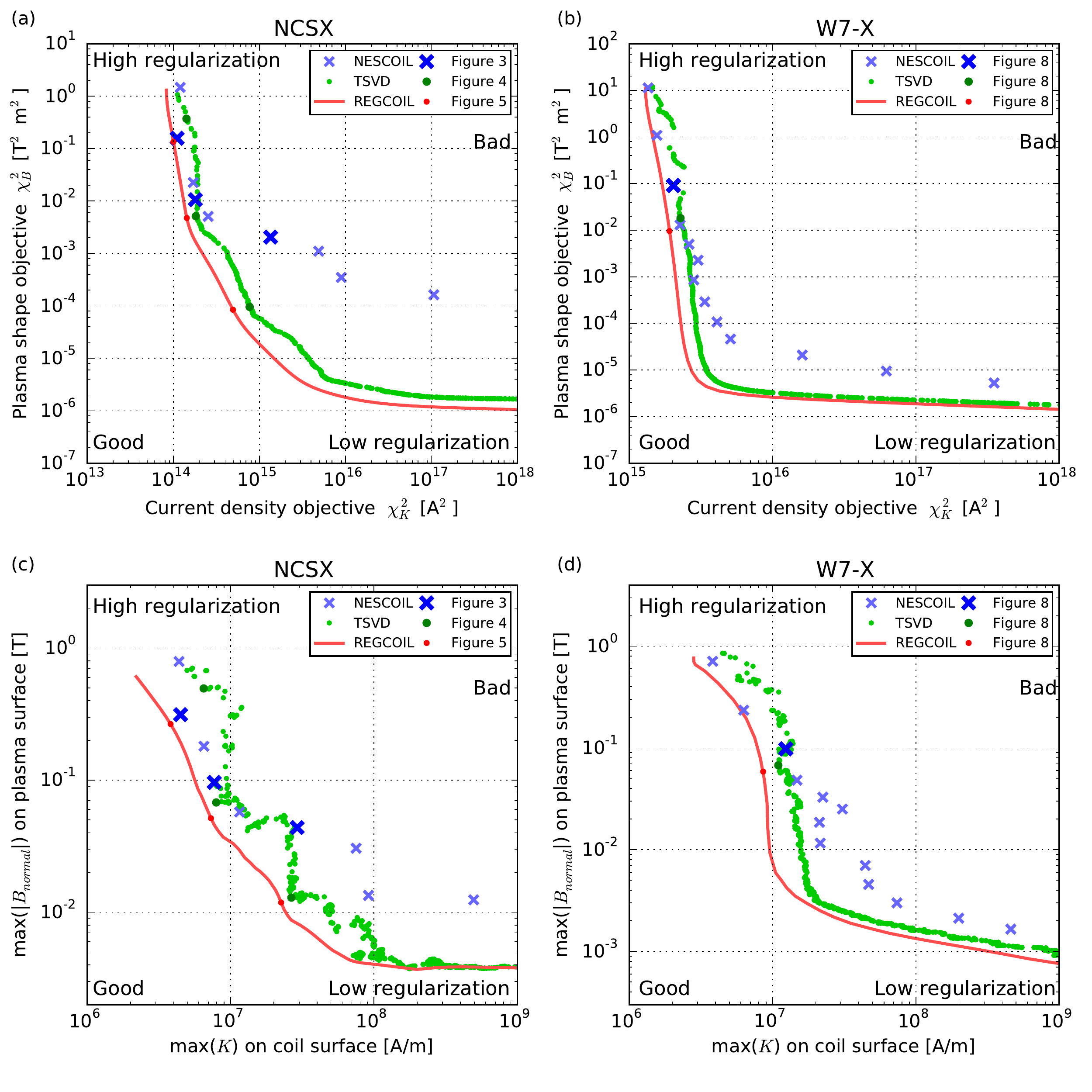}
\caption{(Color online)
Tradeoffs in the competing criteria of $\Bnormal$ versus current density $K$ for the three algorithms and two geometries.
As the level of regularization is scanned for each algorithm, a series of points is obtained if the regularization is discrete 
(\nescoil~and TSVD), while a continuous curve is obtained for \regcoil~since the regularization is continuous.
A point lying above and to the right of another point
is `dominated' in the sense that both criteria are simultaneously improved at the latter point. Figures (a)-(b)
show the tradeoff curves for the surface-average-squared quantities. Figure (c)-(d) show
the tradeoff curves for the maximum quantities. Results from \regcoil~consistently dominate
results from the other methods.
\label{fig:Pareto}}
\end{figure}

Comparing figures \ref{fig:decreasingRegularization_nescoil}-\ref{fig:decreasingRegularization_regcoil}, it can be seen
that each row of the \regcoil~results has comparable or better $K$ and $\Bnormal$ to the corresponding rows for the other two methods.
This behavior is systematic, as shown in figure \ref{fig:Pareto}.
Parts (a)-(b) of the figure show the averaged $\Bnormal$ error $\chi^2_B$ and squared current density $\chi^2_K$
for the three algorithms and two geometries, scanning the amount of regularization for each algorithm over a wide range. For \nescoil,
results are displayed for $N_F = 1,2,3,\ldots$.  (Using different limits for the number of poloidal and toroidal 
Fourier modes results in worse performance, i.e. higher $\chi^2_K$ for given $\chi^2_B$ and vice-versa, corresponding to points above and to the right of those plotted.)
For the TSVD method, the level of regularization is scanned by varying the number of singular values and vectors retained in
the pseudoinverse from $1,2,3,\ldots$. For \regcoil, $\lambda$ is scanned from 0 to $\infty$.
When $\lambda \to \infty$, corresponding to minimizing $\chi^2_K$ with no contribution from $\chi^2_B$, a finite and regular solution results
for the current potential, corresponding to current filaments separated by the maximum possible average distance.
When $\lambda \to 0$, corresponding to minimizing $\chi^2_B$ with no contribution from $\chi^2_K$, the current potential diverges
as numerical resolution is increased, just as for \nescoil.  
The solutions displayed in figures \ref{fig:decreasingRegularization_nescoil}-\ref{fig:decreasingRegularization_regcoil} are
indicated with bold symbols.
Unsurprisingly, \regcoil~consistently obtains
the best (lowest) $\chi^2_B$ for given $\chi^2_K$, and lowest $\chi^2_K$ for given $\chi^2_B$,
since the \regcoil~method was constructed precisely to compute this optimum. 
In the terminology of multi-objective optimization, the \nescoil~and TSVD solutions are `dominated' by those
of \regcoil, corresponding to the former being above and to the right of the latter.
Indeed, since $\chi^2_B$ and $\chi^2_K$ are convex functions of $\currentPot_j$, minimization of the weighted sum
(\ref{eq:chi2tot}) directly yields
the so-called `Pareto frontier' of all possible non-dominated solutions
to the multi-objective optimization problem \cite{steuer}, so no other algorithm could possibly obtain lower $\chi^2_B$ and $\chi^2_K$ simultaneously.
The TSVD method tends to perform a bit worse than \nescoil~for high amounts of regularization, but performs
better than \nescoil~for low levels of regularization.

Next, figures \ref{fig:Pareto}.c-d show the tradeoff in maximum rather than surface-average-squared $|\Bnormal|$ and $K$.
The \regcoil~method is not directly minimizing these criteria,
there is no fundamental reason why the \regcoil~method must always outperform
\nescoil~and the TSVD methods by these measures. Nonetheless, the figures show that the \nescoil~and TSVD results
lie above and to the right of the \regcoil~results, aside from a few NCSX TSVD solutions at very low regularization on the far right.
(The current potential at this low level of regularization is highly contorted so these solutions are uninteresting.)
Evidently, the fact that \regcoil~ minimizes
the surface-average-squared quantities means it tends to do a good job minimizing the maximum quantities,
even though it is not a direct optimization for the latter.

One way to understand the improved performance of \regcoil~over \nescoil~is the following.
In \nescoil, one only optimizes the current potential  within an artificially circumscribed space,
the space of functions that have a very short Fourier expansion.
By contrast, in the \regcoil~approach, the current potential can be optimized over a larger space
of functions, with longer Fourier expansions. 
Although some current potentials with long Fourier expansions correspond to impractical coils,
some current potentials with moderately long Fourier expansions do correspond to practical coils.
By removing the constraint on the Fourier series length,
both $\chi^2_B$ and $\chi^2_K$ can be improved simultaneously.
Since the Fourier series length is a less physically meaningful criterion than $\chi^2_K$ (which corresponds to the inverse distance between coils),
it is advantageous to sacrifice the former for the latter.

Another noteworthy observation is that for a given residual $\Bnormal$,
the TSVD method tends to produce the most contorted
coil shapes. This behavior is particularly visible in the top-right plots of figures
\ref{fig:decreasingRegularization_nescoil}-\ref{fig:decreasingRegularization_regcoil},
and it is related to the trend in figure \ref{fig:Pareto} that the TSVD approach systematically yields higher $\chi^2_K$ than \regcoil~for
given $\chi^2_B$. These aspects of the TSVD method can be understood in light of observations in  \cite{LandremanBoozer}.
In that work, it was pointed out that the SVD does not monotonically order structures in the current potential
from largest to smallest scale, because $\Bnormal$ is driven by the \emph{gradient} of the current
potential, and the gradient emphasizes small-scale structures. While the large-scale structures in the current
potential generate magnetic fields that propagate efficiently from the coil to plasma surface,
it is the small-scale structures in the current potential that most efficiently generate magnetic field locally.
Hence, the TSVD emphasizes small-scale structures, yielding unnecessarily contorted coil shapes.
This problem could possibly be avoided by using a truncated \emph{generalized} SVD \cite{HansenBook},
using a second matrix related to $\chi^2_K$.

Ideally, one would like to have an automated procedure to choose an appropriate level of regularization.
Such automation would be particularly useful for the `stage 1' fixed-boundary plasma optimization, 
to identify plasma shapes that can be produced by distant coils,
since the current potential may then need to be computed thousands of times.
In other fields, one popular method for automatically choosing the level of regularization for an ill-posed problem
is the `L-curve' method \cite{Hansen1992,HansenBook}. This method is based on the observation that for a wide class of regularized ill-posed problems,
a log-log plot of the two competing criteria typically has a characteristic L shape, as can be seen in
figure \ref{fig:Pareto}.b.  In the L-curve method, one adopts the heuristic that the regularization level should be chosen
to obtain a solution on the corner of this L-shaped curve. Unfortunately, the L-curve method appears to be unhelpful
for the present problem of computing the current potential for stellarator coils, for two reasons.
First, 
while an L-shaped curve does exist for W7-X (figure \ref{fig:Pareto}.b),
 the appropriate level of regularization for reasonable equilibrium reconstruction
and reasonable coil curvature actually corresponds to a point on the vertical part of the L,
such as the bold symbols in figure \ref{fig:Pareto}.b.
(This will be demonstrated in the next section.)
Second, for the NCSX case, the L curve in figure \ref{fig:Pareto}.a is quite broad, so it is hard to identify
a unique corner.
Instead of the L-curve heuristic, an automatic procedure for choosing the level of regularization
could be to find the point on the tradeoff curve (figure \ref{fig:Pareto}) at which some engineering limit 
is reached,
such as the maximum allowable current density $K$.

\section{Free boundary reconstructions using discrete coils}

\begin{figure}[h!]
\includegraphics[width=3in]{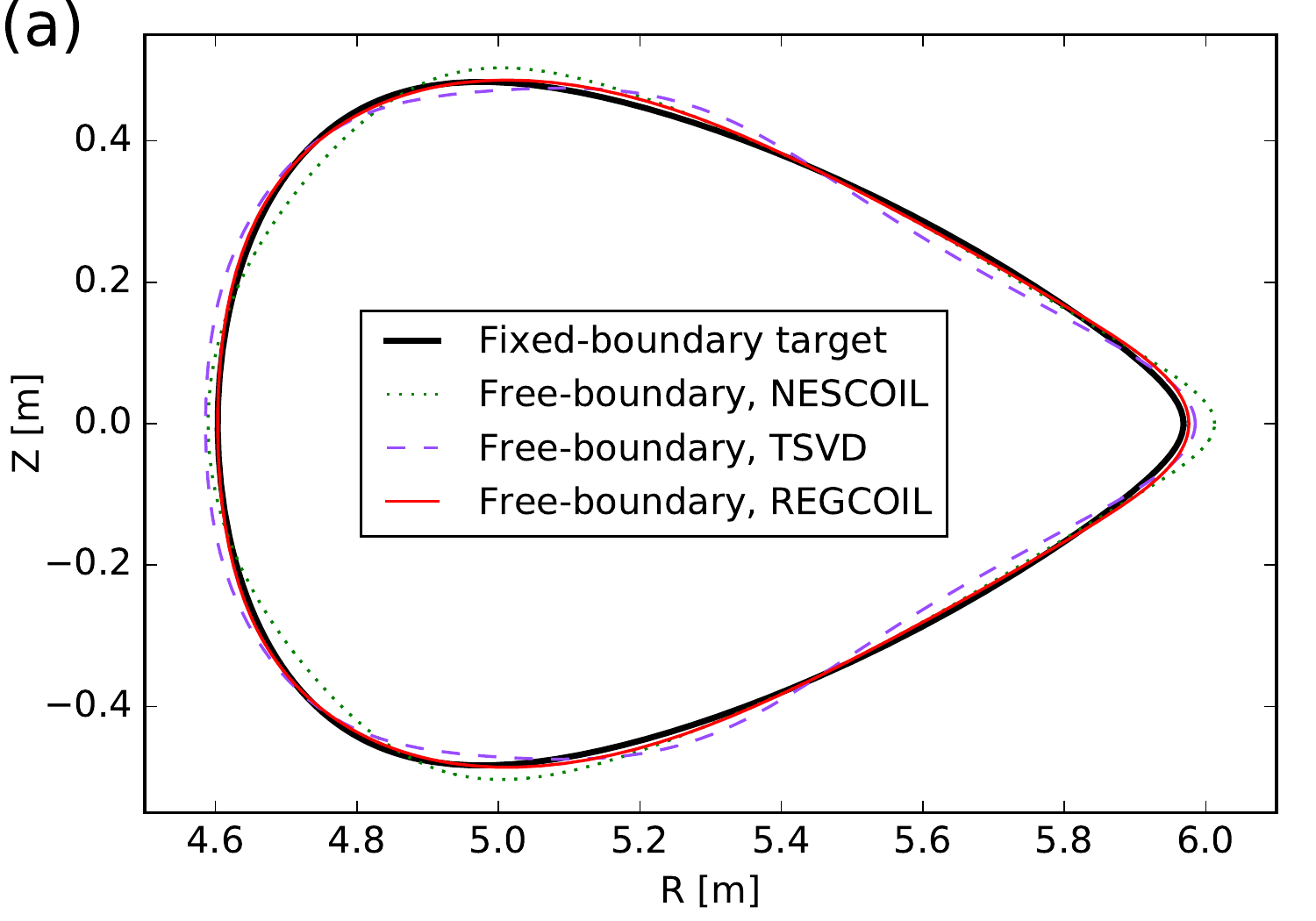}
\includegraphics[width=3in]{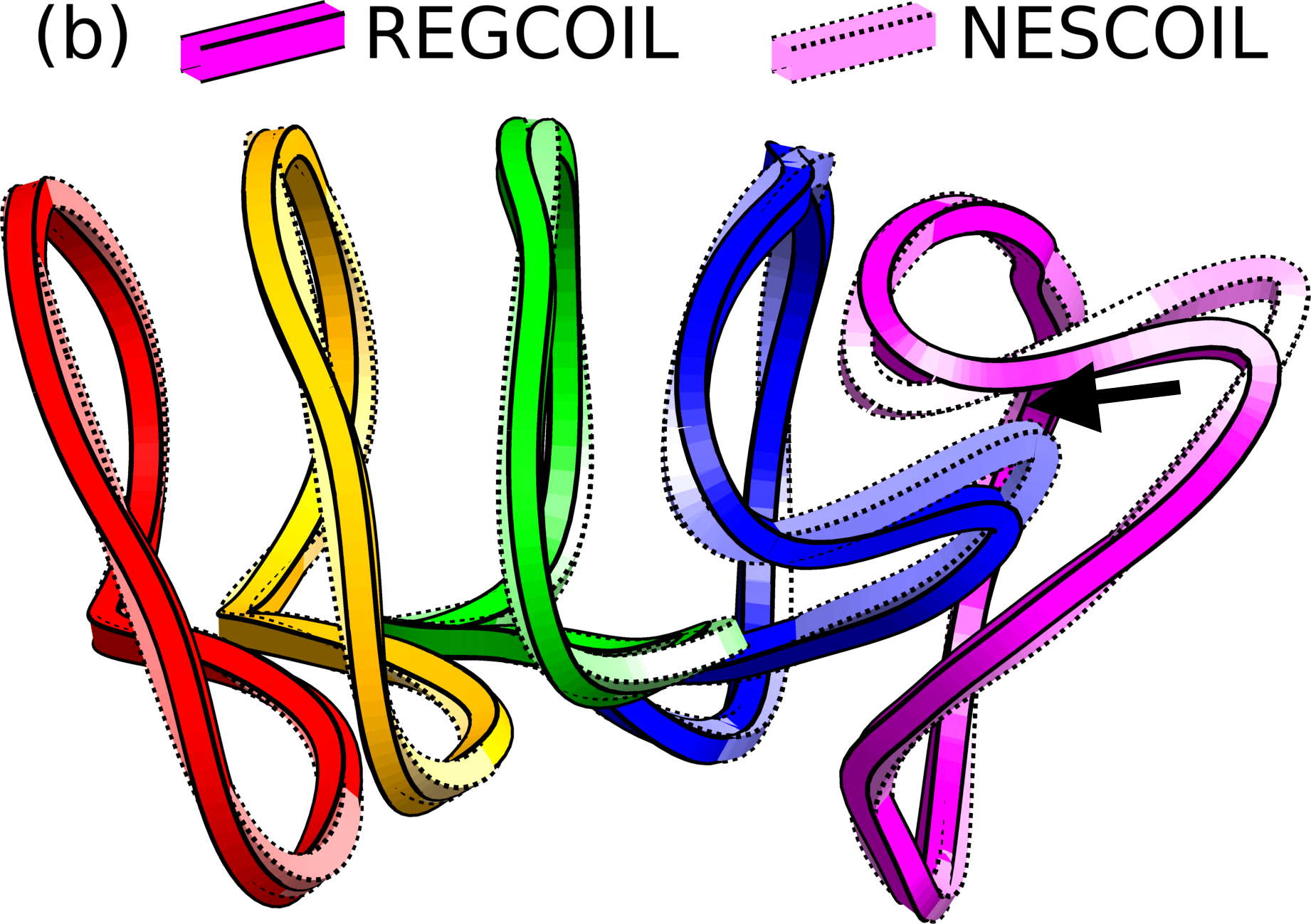}
\includegraphics[width=3in]{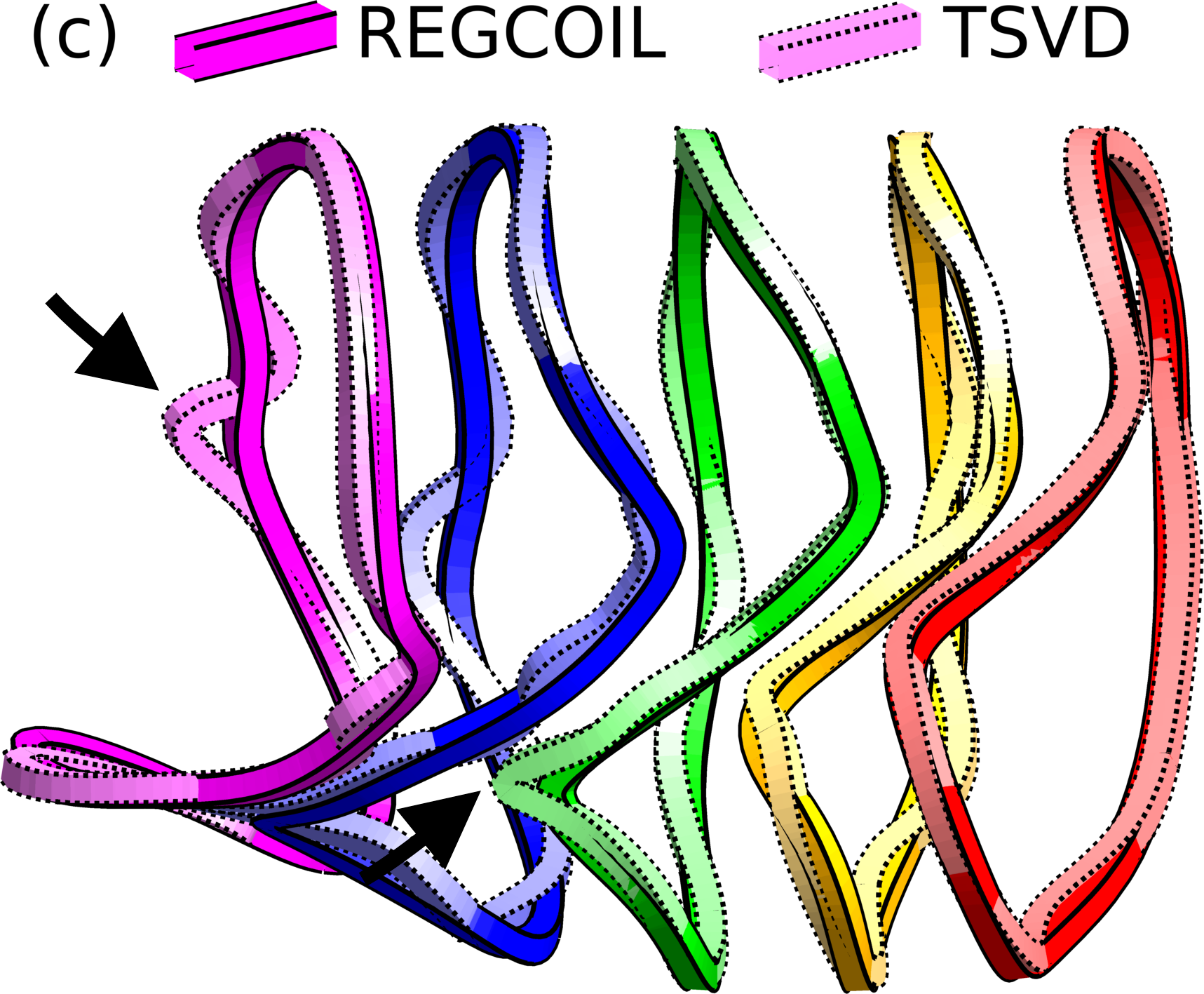}
\caption{(Color online)
Compared to the TSVD and \nescoil~methods,
\regcoil~can achieve superior reconstruction
of the target plasma shape and superior coil shapes simultaneously. 
Here, coilsets for W7-X were computed using the three methods, with an appropriate level of regularization
for a reasonable reconstruction of the target shape. The fixed-boundary target shape is shown in (a)
along with free-boundary \vmec~reconstructions using the three methods with sets of 50 modular coils.
The \regcoil~reconstruction is closest to the target shape, associated with the lower $\chi^2_B$ and lower $\max(|\Bnormal|)$
in figure \ref{fig:Pareto}.
The 5 unique coil shapes from \nescoil~and \regcoil~for this reconstruction are shown overlaid in (b).
More space is available between the blue and magenta coils from \regcoil,
associated with lower $\chi^2_K$ in figure \ref{fig:Pareto}.
The TSVD and \regcoil~coilsets used in (a) are shown overlaid in (c),
where the TSVD coils display regions of unnecessarily high curvature.
The \regcoil~coils in (b) and (c) are identical, although the viewing angle is different.
\label{fig:freeb_W7X}}
\end{figure}

Another demonstration of the superiority of \regcoil~is shown in figure \ref{fig:freeb_W7X}.
Here we use the W7-X geometry from figure \ref{fig:geometry}.c-d,
and we compute sets of 50 discrete coils (5 per half period, as for the real experiment)
using each of the three algorithms. 
For these calculations, the regularization parameter for each method was chosen to select points relatively
close to each other in figures \ref{fig:Pareto}.b and d., shown in bold symbols, each suitable
for a reasonable quality of plasma shape reconstruction.
Then \vmec~is run in free-boundary mode \cite{VMEC1986} to compute an equilibrium supported
by each coilset. Figure \ref{fig:freeb_W7X}.a shows that while all three coilsets do
a reasonable job producing the desired plasma shape, the coilset from \regcoil~does slightly better than the others.
The coilsets from \nescoil~and \regcoil~are overlaid in figure \ref{fig:freeb_W7X}.b. It can be seen
that for the \regcoil~set, there is more space between the blue and magenta coils,
associated with a reduction in the maximum $K$.
In figure \ref{fig:freeb_W7X}.c the coilsets from the TSVD method and \regcoil~are overlaid.
It can be seen that the coils from the TSVD method have several regions with significantly higher
curvature than the corresponding regions of the \regcoil~coils, particularly in the magenta
and green coils.
Neglecting the coil thickness, the distance of closest approach between the blue and magenta coils from \regcoil~is 0.280 m, larger (better)
than the corresponding distances 0.233 m for \nescoil~and 0.174 m for TSVD.
Similarly, the separation between consecutive magenta coils in adjacent half-periods is 0.331 m for \regcoil,
much improved from 0.141 m for \nescoil~and 0.199 m for TSVD.
Thus, the coilsets generated from \regcoil~simultaneously do a better job of producing the desired plasma shape
and maximizing space between the coils for ports and other components.
This superiority is reflected in the fact that the \regcoil~solution used in
figure \ref{fig:freeb_W7X} has lower $\chi^2_B$, lower $\chi^2_K$,
lower $\max(|\Bnormal|)$, and lower $\max(K)$ than the solutions from the other two methods.

\begin{figure}[h!]
\includegraphics[width=3in]{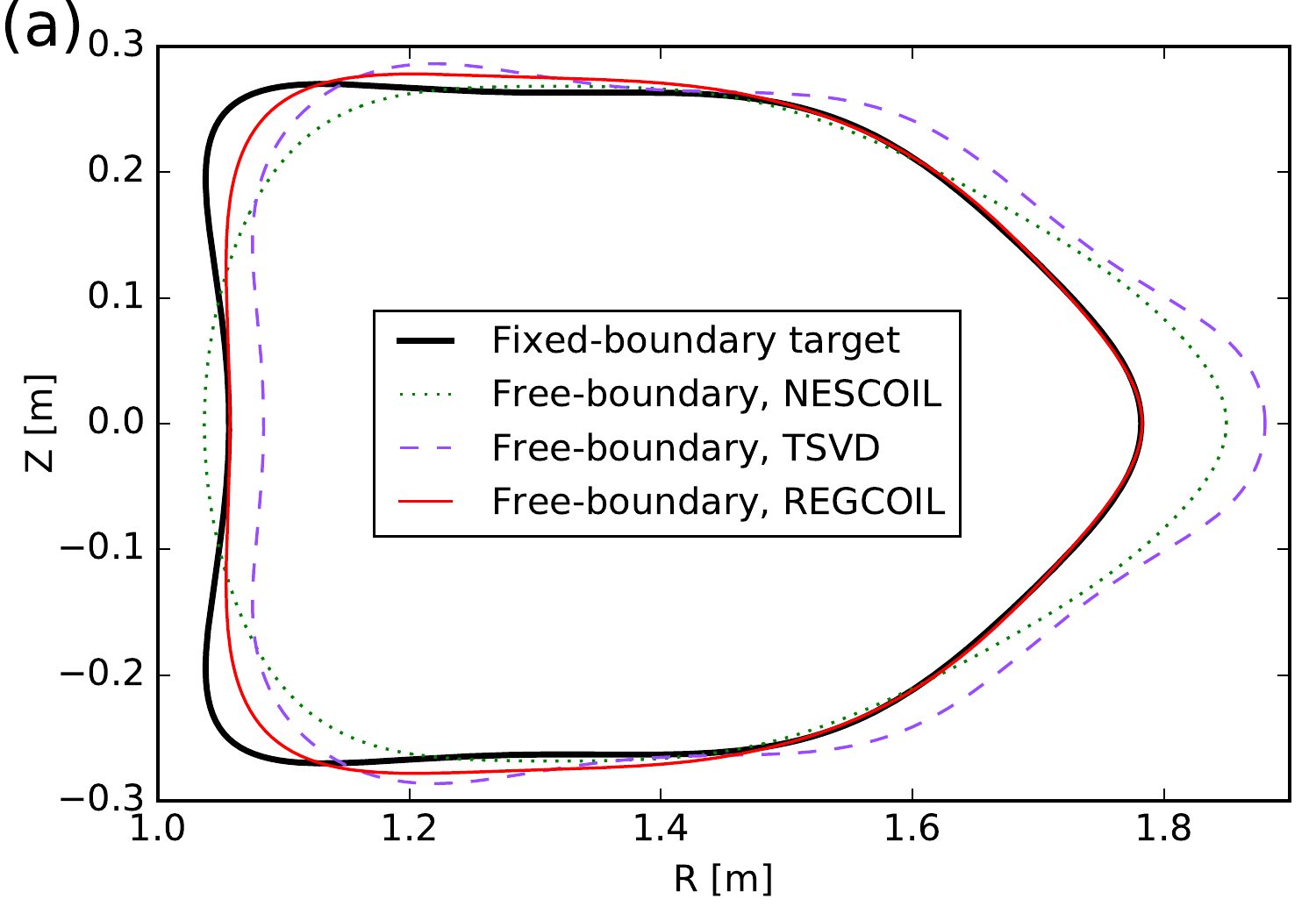}
\includegraphics[width=3in]{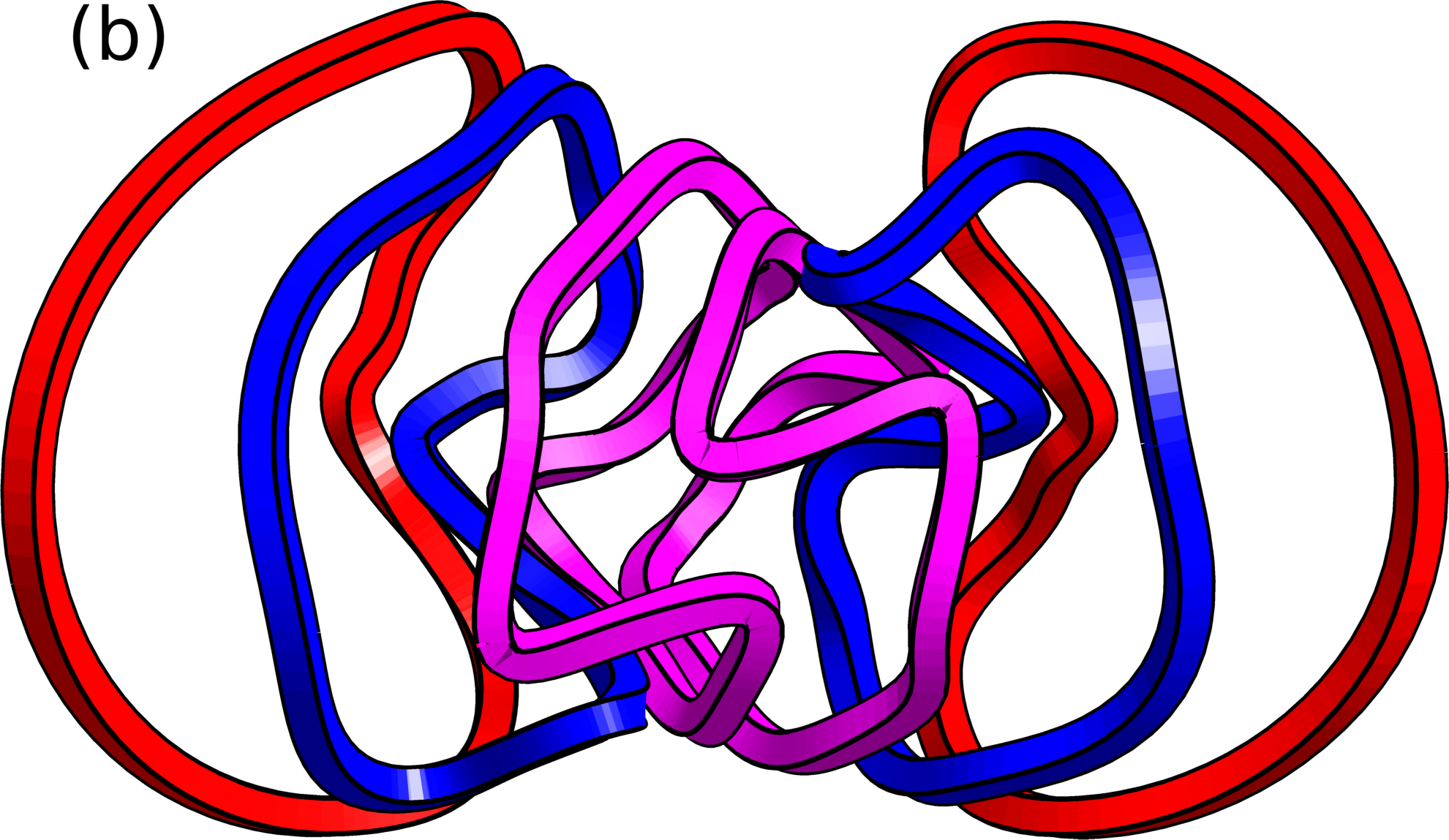}
\caption{(Color online)
Free-boundary NCSX reconstruction using 3 coils per half period.
(a) The best possible reconstruction without saddle coils from \nescoil~and the TSVD method
is not nearly as accurate as with a coilset from \regcoil.
Two half-periods of the \regcoil~coilset used for (a) are shown in (b).
\label{fig:freeb_NCSX}}
\end{figure}

Another practical advantage of \regcoil~is that for a given $\chi^2_B$, 
modular coils are less likely to bifurcate into saddle coils, indicated by contours of the current potential that do not link the plasma.
This trend can be seen by comparing the middle rows of figures
\ref{fig:decreasingRegularization_nescoil}-\ref{fig:decreasingRegularization_regcoil}.
In a calculation of non-planar modular coils such as this,
the saddle coils are undesirable since they represent additional components to fabricate
and they block port access.
If one does not want saddle coils, the ability of \nescoil~and the TSVD method
to reconstruct the NCSX target plasma shape is quite limited.
This limitation is shown in the free-boundary reconstruction for NCSX in figure \ref{fig:freeb_NCSX}.a,
using 3 coils per half-period. (For simplicity we take the coils to lie on equally spaced contours, carrying equal current.)
For the \nescoil~and TSVD methods, we choose the regularization parameters to
give the most accurate reconstruction of the target plasma
shape possible without saddle coils. In other words, we start with the maximum possible
regularization, then lower the regularization
until the topology of the three relevant contours is about to change.
Slightly higher regularization is used for \regcoil~to smooth the coil shapes a bit.
Figure \ref{fig:freeb_NCSX} demonstrates that while none of the algorithms
is able to achieve a perfect reconstruction of the target plasma shape, \regcoil~achieves
a much better reconstruction than the other methods. The \regcoil-derived coilset used for
 \ref{fig:freeb_NCSX}.a is shown in figure \ref{fig:freeb_NCSX}.b

\section{Coordinate dependence of the methods}

Since the coordinates $\theta'$ and $\zeta'$ parameterizing the coil surface appear explicitly in \nescoil's regularization scheme
(the Fourier series truncation),
the resulting coil shapes are not the same if the coil surface is reparameterized using some other equally valid angle coordinates.
This property is undesirable for several reasons.
It places an extra demand on the code user,
as he or she must be careful to use `good' coordinates to parameterize the coil surface,
and it is not actually clear what the best coordinates are.
Since \nescoil~may well return superior coil shapes if the surface is reparameterized,
it is never clear if the best solution to the optimization problem has been found.
Another problem arises if \nescoil~is called within fixed-boundary plasma optimization in order to find plasma shapes consistent with good coils, as in \cite{pomphrey}.
In such an optimization, a large residual $\chi^2_B$ or high coil complexity from \nescoil-derived coils may not really mean that the plasma
shape is hard to produce from external coils; it may instead indicate that the automatically-generated coil surface
happens to be parameterized with coordinates for which \nescoil~performs suboptimally.
Hence the fixed-boundary plasma optimizer may unfairly penalize such plasma shapes.

By contrast, the $\chi^2_K$ term used for regularization in \regcoil~is defined by a coordinate-independent expression,
and hence the coil shapes resulting from \regcoil~are independent of the coordinates used.

The TSVD solver implemented in the \nescoil~code \cite{valanjuPoster,pomphrey} and described here yields coordinate-dependent results, just as for standard
\nescoil. However as discussed in \cite{LandremanBoozer}, the TSVD method can be reformulated to yield coordinate-independent results.
The central idea is to replace the Fourier modes with basis functions that are defined to be orthogonal
under the coordinate-independent operation $\int d^2a (\ldots)$.

\begin{figure}[h!]
\includegraphics[width=3in]{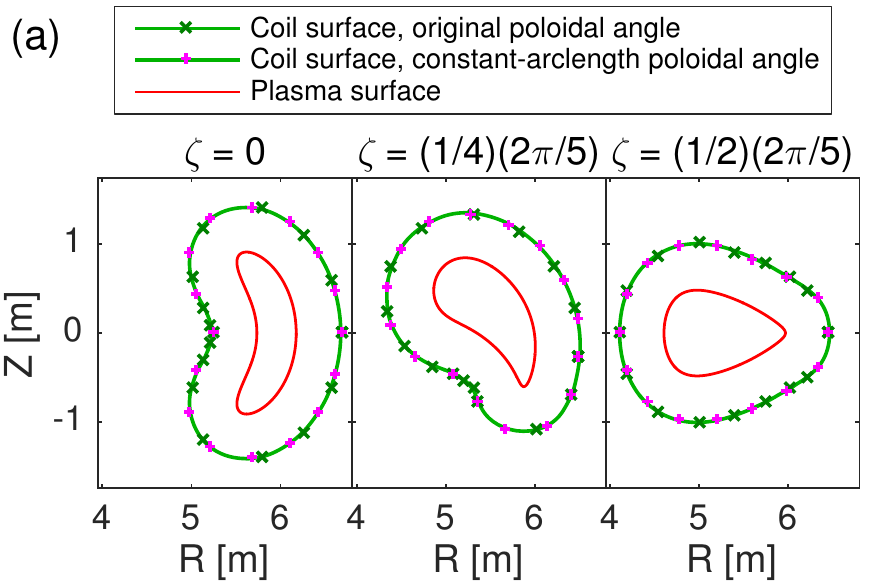}
\includegraphics[width=3in]{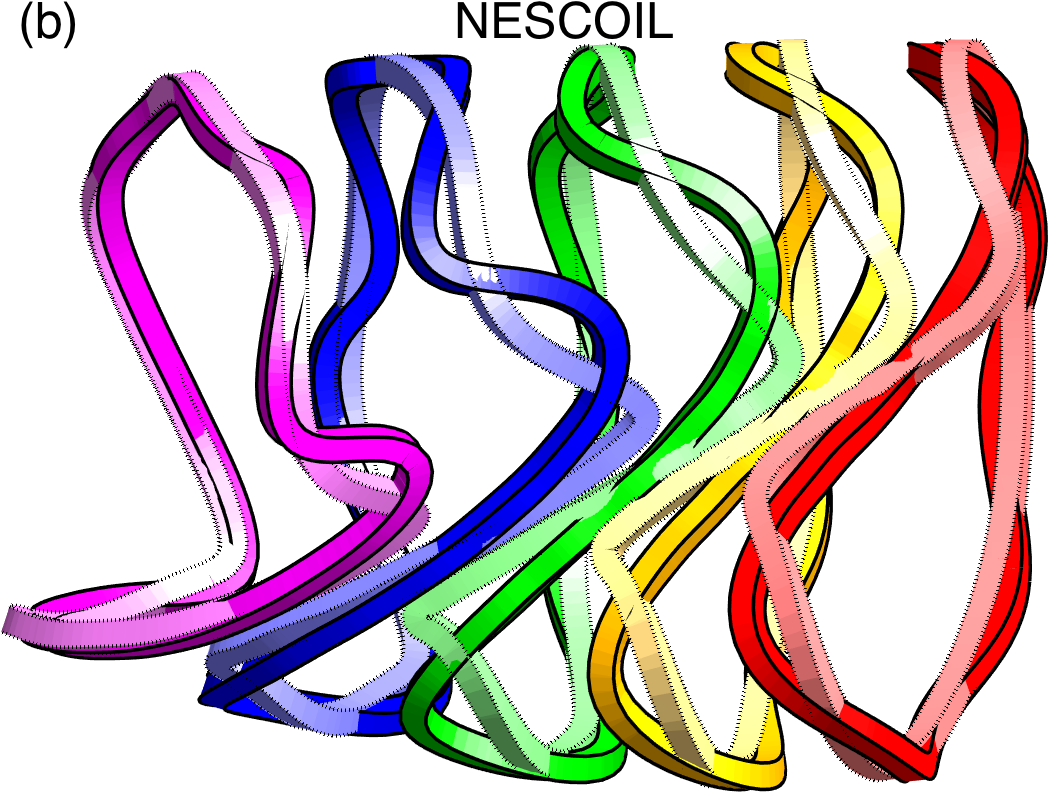}
\includegraphics[width=3in]{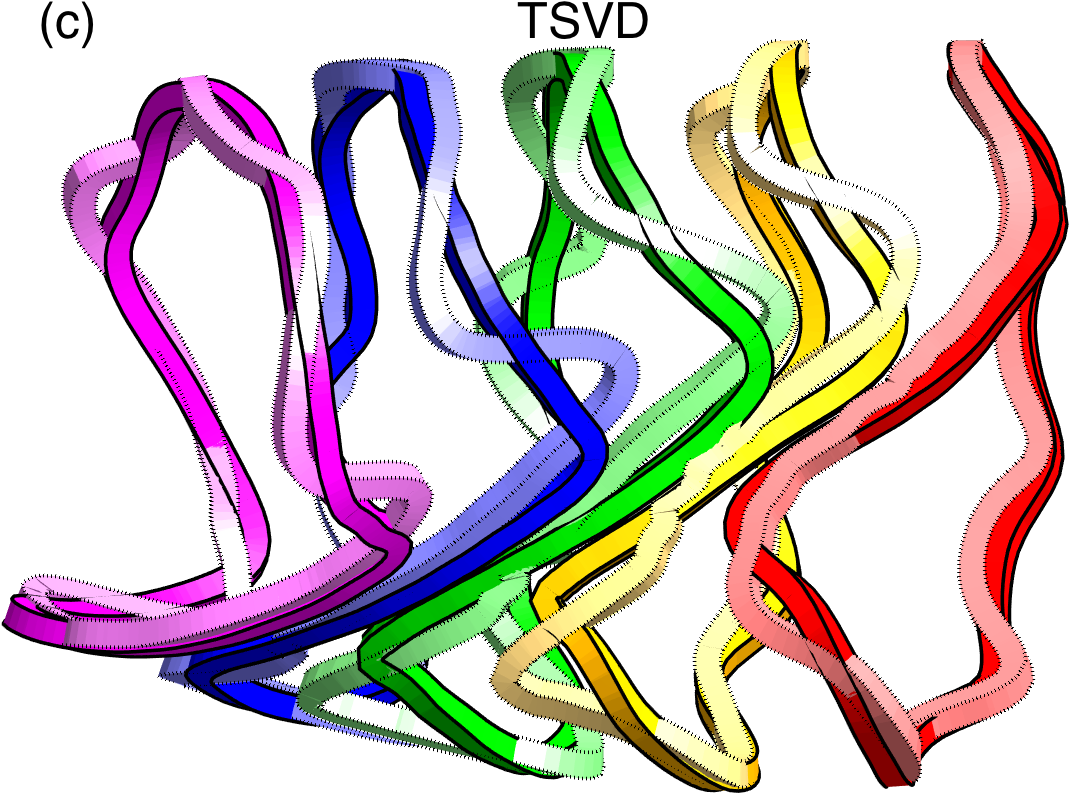}
\includegraphics[width=3in]{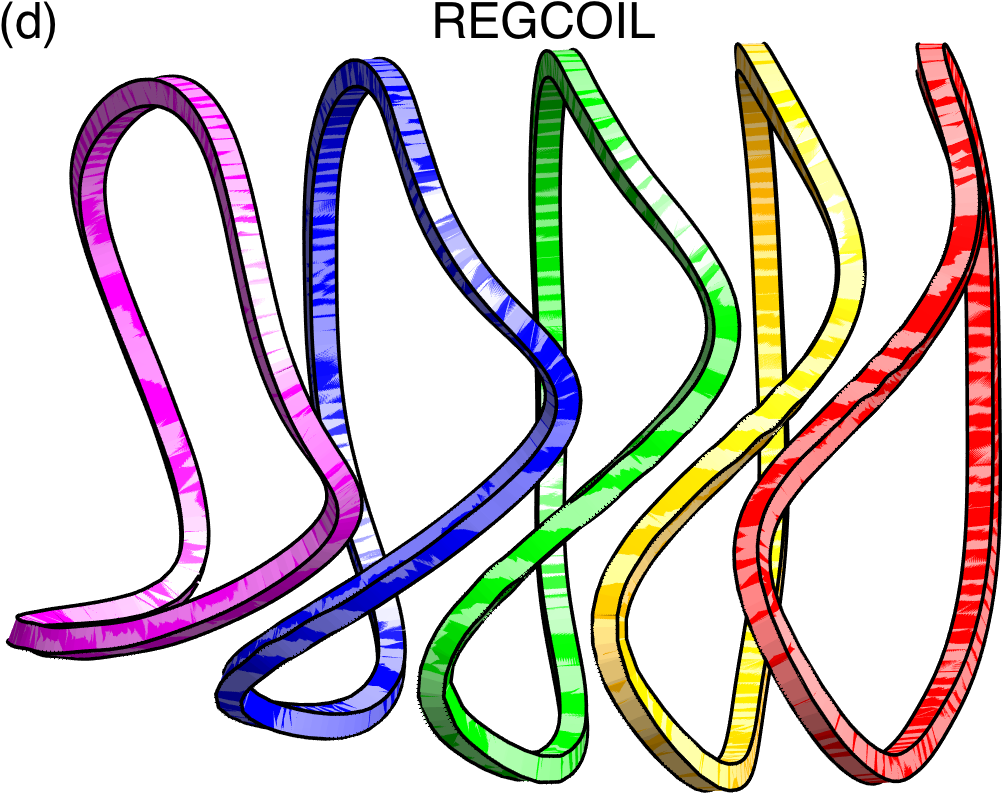}
\caption{(Color online)
Disturbingly, if the coil surface is reparameterized using a different poloidal angle as in (a), the
coil shapes computed by (b) the original \nescoil~method and (c) the TSVD solver can change significantly. Shown here are the 5 independent W7-X coil shapes computed for a uniform 50cm offset from the plasma surface. Pale colors with dotted edges indicate the result using the original poloidal angle from the offset-surface generation algorithm; dark colors with solid edges indicate the result if the same coil surface is re-parameterized using a constant-arclength poloidal angle. 
(d) This pathology is not present in the \regcoil~approach, where the coils computed using the different surface parameterizations are identical (within a tiny discretization error that 
causes the striped pattern upon rendering.)
\label{fig:coordinateDependenceOfNescoil}}
\end{figure}

These concepts are illustrated in figure \ref{fig:coordinateDependenceOfNescoil}.
We first generate a coil surface with uniform offset of 50 cm from the W7-X plasma surface using the algorithm 
of appendix \ref{sec:offsetAlgorithm},
(used in the {\ttfamily BNORM} code often run to prepare \nescoil~input files).
This algorithm imposes a particular choice of poloidal angle on the coil surface. The coil surface can then be reparameterized using
another poloidal angle, such as the angle with constant arclength $|\partial \vect{r}'/\partial \theta'|$ at each $\zeta'$.
Uniformly spaced points in these two poloidal angles are shown by the $+$ and $\times$ symbols in figure \ref{fig:coordinateDependenceOfNescoil}.a. Figure \ref{fig:coordinateDependenceOfNescoil}.b
shows the \nescoil~results for the two surface parameterizations, using Fourier modes $|m| \le 4$ and $|n| \le 4$.
The resulting 5 independent coil shapes (assuming 5 coils per half-period)
for the two different surface parameterizations are plotted on top of each other to show the differences.
Figure \ref{fig:coordinateDependenceOfNescoil}.c shows the TSVD solutions for the two surface parameterizations, retaining 125 singular values.
Figure \ref{fig:coordinateDependenceOfNescoil}.d shows the \regcoil~solutions for the two surface parameterizations.
Evidently, the coil shapes computed by \nescoil~and the TSVD approaches are substantially different when these two 
reasonable parameterizations of the same coil surface are used. In contrast the \regcoil~results are universal, independent
of the surface parameterization.

Parameterization-dependent behavior as in \nescoil~could also occur in nonlinear coil optimization codes, in which the coil shapes are directly optimized \cite{onset,coilopt,Brown}. 
In these codes, one should ensure the dominant source of regularization is provided by a physical coordinate-independent quantity
like $\chi^2_K$ or the coil radius of curvature, not by the coordinate-dependent spline or Fourier representation
of the coil shape.

\section{Conclusions}
\label{sec:conclusions}

In this work we have compared three methods for computing the shape of stellarator coils: the classical \nescoil~procedure \cite{merkel_nescoil}, TSVD solution of the \nescoil~least-squares problem \cite{valanjuPoster,pomphrey}, and the Tikhonov regularization method we have called \regcoil.
The problem of determining the external currents that produce a given plasma shape
is fundamentally ill-posed, requiring some form of regularization to obtain a meaningful and numerically well-behaved solution,
and the three methods above each employ a different regularization strategy.
All three methods have the advantage that they are guaranteed to provide a unique solution (for given choice of regularization parameter and coil surface parameterization), i.e
there is no chance of getting `stuck' in a local rather than global optimum as could occur in non-convex formulations. Also all three methods are `linear' in the sense that there are no nonlinear equations involved which would require iterative solution. Hence all methods are comparably robust and fast. 
We find the TSVD method unnecessarily emphasizes small-scale structures in the coil shapes,
for the reasons detailed in \cite{LandremanBoozer}.
Moreover, since \regcoil~employs a more physically significant regularization approach than the other methods,
we find that  \regcoil~has several compelling advantages:
\begin{enumerate}
\item It consistently yields lower surface-integrated-squared and maximum $|\Bnormal|$ on the target plasma surface at the same time as lower surface-integrated-squared and maximum current density $K$ on the coil winding surface (figure \ref{fig:Pareto}).
Correspondingly, \regcoil~simultaneously achieves more accurate reconstruction of the desired plasma shape (figure \ref{fig:freeb_W7X}.a and \ref{fig:freeb_NCSX}.a) while increasing the minimum distances between coils (figure \ref{fig:freeb_W7X}.b-c).
Due to this last fact, \regcoil~reduces the chances of coil-coil collisions and improves access for ports and maintenance.
\item \regcoil~allows finer control over the amount of regularization, since the regularization parameter is continuous rather than discrete.
\item Unlike the other methods, \regcoil~yields coil shapes that are independent of the arbitrary choice of coordinates used to parameterize the coil surface (figure \ref{fig:coordinateDependenceOfNescoil}),
so the user need not worry if the surface has been parameterized using `good' angles.
\end{enumerate}
For these reasons, we contend that \regcoil~should be used instead of \nescoil~for applications in
which a fast and robust method for coil calculation is needed, such as when targeting coil complexity in
fixed-boundary plasma optimization, or for scoping of new stellarator geometries.

\begin{acknowledgments}
The author is grateful to Michael Drevlak for providing the W7-X plasma and coil surface data, and to 
Neil Pomphrey for providing the NCSX plasma and coil surface data.
We also acknowledge illuminating conversations about this topic with Allen Boozer and Mike Zarnstorff, and
thank Sam Lazerson for 
providing
the \vmec~and \nescoil~codes.
This work was supported by the
U.S. Department of Energy, Office of Science, Office of Fusion Energy Science,
under Award Number DE-FG02-93ER54197.
Computations were performed on the Edison and Cori systems at
the National Energy Research Scientific Computing Center, a DOE Office of Science User Facility supported by the Office of Science of the U.S. Department of Energy under Contract No. DE-AC02-05CH11231.
\end{acknowledgments}

\appendix

\section{Equations for \nescoil~and \regcoil}
\label{sec:equations}

Here we derive the equations used for the \nescoil~and \regcoil~methods.
First we derive an expression for $\chi^2_B$,
equivalent to the calculation in \cite{merkel_nescoil}.
Next we derive an expression for $\chi^2_K$.
The derivative of the total objective function then yields a linear system
for the single-valued part of the current potential.

We first note that the surface integral of any quantity $Q$ (such as appears in (\ref{eq:chi2B}) and (\ref{eq:chi2K})) can be written
\begin{equation}
\int d^2a\; Q
= \sum_{\ell=1}^{n_p} 
\int_0^{2\pi} d\theta
\int_0^{2\pi/n_p} d\zeta
\; NQ.
\label{eq:area}
\end{equation}
Here $n_p$ is the number of identical toroidal periods (e.g. 5 for W7-X), the toroidal period is indexed by $\ell$, 
$N = |\vect{N}|$, and
\begin{equation}
\vect{N} = \frac{\partial \vect{r}}{\partial \zeta} \times \frac{\partial \vect{r}}{\partial \theta}
\label{eq:N}
\end{equation}
is a non-unit-length surface normal vector.

The Biot-Savart law gives the magnetic field due to a surface current density $\vect{K}$ as
\begin{equation}
\vect{B}(\vect{r}) = \frac{\mu_0}{4\pi} \int d^2a' \frac{\vect{K'} \times (\vect{r}-\vect{r}')}
{| \vect{r}-\vect{r}' |^3}.
\end{equation}
As in the main text, primed coordinates are those on the coil surface, and other primed quantities like $\vect{K}'$ and $\vect{r}'$ are shorthand for $\vect{K}(\theta', \zeta')$ and $\vect{r}(\theta', \zeta')$.
Applying (\ref{eq:N}) and (\ref{eq:area}),
and noting 
\begin{eqnarray}
\vect{N}\times\nabla \theta &=& -\partial \vect{r} / \partial \zeta, \label{eq:identity}\\
\vect{N}\times\nabla \zeta &=& \partial \vect{r} / \partial \theta, \nonumber
\end{eqnarray}
we find
\begin{equation}
\vect{B}(\vect{r}) = \frac{\mu_0}{4\pi} 
\sum_{\ell'=1}^{n_p}
 \int_0^{2\pi}d\theta'
 \int_0^{2\pi/n_p}d\zeta' 
\frac{1}{| \vect{r}-\vect{r}' |^3}
\left( \frac{\partial \vect{r}'}{\partial \theta'} \frac{\partial \currentPot'}{\partial \zeta'}
-\frac{\partial \vect{r}'}{\partial \zeta'} \frac{\partial \currentPot'}{\partial \theta'} \right)
\times (\vect{r}-\vect{r}').
\label{eq:BiotSavart2}
\end{equation}
Considering the normal component of (\ref{eq:BiotSavart2}) on the plasma surface,
the contribution to $\Bnormal$ from the secular terms in (\ref{eq:currentPotDecomp}) is
\begin{eqnarray}
\Bnormal^{GI}
&=& \frac{\mu_0}{8\pi^2} 
\sum_{\ell'=1}^{n_p} 
\int_0^{2\pi}d\theta'
\int_0^{2\pi/n_p}d\zeta' 
\frac{1}{| \vect{r}-\vect{r}' |^3}
\left( G\frac{\partial \vect{r}'}{\partial \theta'} 
-I \frac{\partial \vect{r}'}{\partial \zeta'} \right)
\times (\vect{r}-\vect{r}') \cdot \vect{n} \\
&\approx &
\frac{\mu_0}{8\pi^2} 
\Delta_{\theta'} \Delta_{\zeta'} \sum_{\ell'=1}^{n_p} \sum_{i_\theta'} \sum_{i_\zeta'} 
\frac{1}{| \vect{r}-\vect{r}' |^3}
\left( G\frac{\partial \vect{r}'}{\partial \theta'} 
-I \frac{\partial \vect{r}'}{\partial \zeta'} \right)
\times (\vect{r}-\vect{r}') \cdot \vect{n}.
\label{eq:h}
\end{eqnarray}
In (\ref{eq:h}) we have discretized the $\theta'$ and $\zeta'$ integrals
by introducing uniform grids in these coordinates with spacing $\Delta_{\theta'}$ and $\Delta_{\zeta'}$,
with domains $[0,2\pi)$ and $[0,2\pi/n_p)$,
and indexed by $i_{\theta}'$ and $i_{\zeta}'$.
Equation (\ref{eq:h}) corresponds to $h/N$ in the notation of \cite{merkel_nescoil}.

Next, we integrate (\ref{eq:BiotSavart2}) by parts in $\theta'$ and $\zeta'$ to
remove the derivatives on $\currentPot'$, and we take the component normal to the plasma
surface. Thus, we find the normal magnetic field due the single-valued current potential is, after some algebra,
\begin{equation}
\BnormalSV \{ \currentPotSV \}
= \frac{\mu_0}{4\pi N} 
\sum_{\ell'=1}^{n_p} 
\int_0^{2\pi}d\theta'
\int_0^{2\pi/n_p}d\zeta' 
\;\currentPotSV'
\left(
\frac{1}{| \vect{r}-\vect{r}' |^3} \vect{N}\cdot\vect{N}'
-\frac{3}{| \vect{r}-\vect{r}' |^5} [\vect{r}-\vect{r}']\cdot \vect{N} [\vect{r}-\vect{r}']\cdot\vect{N}'
\right).
\end{equation}
Considering the Fourier expansion (\ref{eq:Fourier}), and discretizing
the $\theta'$ and $\zeta'$ integrals as in (\ref{eq:h}), we find
the normal magnetic field driven by the single-valued current potential is
\begin{equation}
\BnormalSV \{ \currentPotSV \}
=\frac{1}{N} \sum_j \currentPot_j g_j,
\end{equation}
where
\begin{equation}
g_j(\theta,\zeta) = \Delta_{\theta'} \Delta_{\zeta'} \sum_{i_\theta'} \sum_{i_\zeta'}
\sincos_j (m_j\theta' - n_j \zeta')\; g(\theta,\zeta,\theta'_{i_\theta'},\zeta'_{i_\zeta'}) \label{eq:discrete_g}
\end{equation}
and
\begin{equation}
g(\theta,\zeta,\theta',\zeta') = \frac{\mu_0}{4\pi} 
\sum_{\ell'=1}^{n_p} \left(
\frac{1}{| \vect{r}-\vect{r}' |^3} \vect{N}\cdot\vect{N}'
-\frac{3}{| \vect{r}-\vect{r}' |^5} [\vect{r}-\vect{r}']\cdot \vect{N} [\vect{r}-\vect{r}']\cdot\vect{N}'
\right)
\end{equation}
are quantities introduced in \cite{merkel_nescoil}.
Using (\ref{eq:chi2B}) and (\ref{eq:area}), the \nescoil~objective function can now be written
\begin{equation}
\chi^2_B =
n_p \Delta_\theta \Delta_\zeta \sum_{i_\theta} \sum_{i_\zeta}
\left[
\sqrt{N}
(\Bnplasma + \Bnexternal + \Bnormal^{GI})
+ \sum_j \currentPot_j \frac{g_j}{\sqrt{N}}
\right]^2.
\label{eq:lsq}
\end{equation}
The $\theta$ and $\zeta$ integrals have been discretized as in (\ref{eq:h}).
Minimization of (\ref{eq:lsq}) has the form of a linear least-squares problem,
$|Ax-b|^2$,
where the matrix $A$ corresponds to $g_j/\sqrt{N}$, the vector $b$ corresponds to $-\sqrt{N} (\Bnplasma + \Bnexternal + \Bnormal^{GI})$,
and the vector $x$ corresponds to $\currentPot_j$.
As with any such problem, it may be solved by the normal equations, $QR$ decomposition, or SVD.
The normal equations approach is typically slightly faster than the other methods but is more affected by roundoff error.

Next, we derive an expression for the current density objective $\chi^2_K$.
Combining (\ref{eq:currentPot}), 
(\ref{eq:currentPotDecomp}), (\ref{eq:N}), and (\ref{eq:identity}),
we find the current density can be written
\begin{equation}
\vect{K}' = \frac{1}{N'} \left( \vect{d} - \sum_j \currentPot_j \vect{f}_j \right),
\end{equation}
where
\begin{equation}
\vect{d} = \frac{G}{2\pi} \frac{\partial \vect{r}'}{\partial \theta'} - \frac{I}{2\pi} \frac{\partial \vect{r}'}{\partial \zeta'}
\end{equation}
is the contribution from the secular terms in the current potential,
and
\begin{equation}
\vect{f}_j = \left[ m_j \frac{\partial \vect{r}'}{\partial \zeta'} + n_j \frac{\partial \vect{r}'}{\partial \theta'} \right]
\cossin_j (m_j\theta' - n_j \zeta') 
\end{equation}
is the contribution from the $j$th mode of the periodic current potential.
Substituting these expressions into (\ref{eq:chi2K}) with (\ref{eq:area}), and discretizing the 
$\theta'$ and $\zeta'$ integrals in the same way as in (\ref{eq:h}),
we obtain
\begin{equation}
\chi^2_K = 
n_p \Delta_{\theta'} \Delta_{\zeta'} \sum_{i_\theta'} \sum_{i_\zeta'} \frac{1}{N'}
\left( \vect{d} - \sum_j \currentPot_j \vect{f}_j \right)^2.
\end{equation}
If one did not care about $\chi^2_B$ and only wanted to
minimize $\chi^2_K$, one would form $\partial \chi^2_K/\partial \currentPot_j=0$ to obtain
the normal equations
\begin{equation}
\sum_k
A^K_{j,k} \currentPot_k = b^K_j,
\end{equation}
with matrix 
\begin{equation}
A^K_{j,k} =
\Delta_{\theta'} \Delta_{\zeta'} \sum_{i_\theta'} \sum_{i_\zeta'} \frac{\vect{f}_j \cdot \vect{f}_k}{N'},
\end{equation}
and  right-hand side 
\begin{equation}
b^K_{j} =
\Delta_{\theta'} \Delta_{\zeta'} \sum_{i_\theta'} \sum_{i_\zeta'} \frac{\vect{d} \cdot \vect{f}_j}{N'}.
\end{equation}

For the combined objective function (\ref{eq:chi2tot}), setting 
$0 = \partial \chi^2/\partial \currentPot_j
= \partial \chi^2_B/\partial \currentPot_j + \lambda \partial \chi^2_K/\partial \currentPot_j$,
we obtain the linear system to solve for \regcoil:
\begin{equation}
\sum_k
A_{j,k} \currentPot_k = b_j,
\end{equation}
with matrix 
\begin{equation}
A_{j,k} = A^B_{j,k}  + \lambda A^K_{j,k},
\end{equation}
\begin{equation}
A_{j,k}^B = \Delta_{\theta} \Delta{\zeta} \sum_{i_{\theta}} \sum_{i_{\zeta}} \frac{g_j g_k}{N},
\end{equation}
and right-hand side
\begin{equation}
b_{j} =  b^B_{j}  + \lambda b^K_{j},
\end{equation}
\begin{equation}
b^B_j = - \Delta_{\theta} \Delta_{\zeta} \sum_{i_{\theta}} \sum_{i_{\zeta}} \left( B_{\mathrm{normal}}^{\mathrm{plasma}} + B_{\mathrm{normal}}^{\mathrm{external}} + G_{\mathrm{normal}}^{GI} \right) g_j
\end{equation}

\section{Offset surface algorithm}
\label{sec:offsetAlgorithm}

Suppose we know a toroidal surface $\vect{r}_p(\theta,\zeta)$,
parameterized by poloidal and toroidal angles $(\theta,\zeta)$,
and we wish to compute a toroidal surface $\vect{r}_c(\theta',\zeta')$ which is offset from $\vect{r}_p$ by a uniform distance $\Delta$.
For each pair of values of the poloidal and toroidal angles $(\theta',\zeta')$
we can obtain $\vect{r}_c(\theta',\zeta')$ as follows. Let $\theta = \theta'$, and solve the 1D nonlinear root-finding problem 
\begin{equation}
\arctan \left( y(\hat\zeta)/x(\hat\zeta) \right) - \zeta'=0
\end{equation}
for the unknown $\hat\zeta$, where $x(\hat\zeta)$ and $y(\hat\zeta)$ are the Cartesian components of the vector $\vect{r}_p(\theta,\hat\zeta) + \vect{n}(\theta,\hat\zeta)  \Delta$.
Here, $\vect{n}$ is the unit normal of the surface $\vect{r}_p$. Thus, $\vect{r}_p(\theta',\hat\zeta)$ is the point
on the original surface such that if we move a distance $\Delta$ in the normal direction, we arrive at the toroidal angle $\zeta'$.
Then the desired  $\vect{r}_c(\theta',\zeta')$ is given by  $\vect{r}_p(\theta,\hat\zeta) +  \vect{n}(\theta,\hat\zeta) \Delta$.
This process can be repeated to generate as many points $(\theta',\zeta')$ on the new surface as desired.
Note that the angle $\zeta'$ parameterizing the new surface is the standard cylindrical angle.

\bibliography{regcoil}

\end{document}